\documentclass[prb,twocolumn,aps]{revtex4-2}
\usepackage{graphicx,amsmath,amssymb,xcolor}
\usepackage{braket,hyperref,lipsum,float}
\usepackage[T1]{fontenc}
\usepackage{silence}
\begin{document}

    \title{A realizable time crystal of four silicon quantum dot qubits}

    \author{Nathan L. Foulk}
    \author{Sankar Das Sarma}
    \affiliation{Condensed Matter Theory Center and Joint Quantum Institute, Department of Physics, University of Maryland, College Park, Maryland 20742-4111 USA}

    \begin{abstract}
        We demonstrate that exciting possible realizations of quantum Floquet matter are within reach for modern silicon spin qubits based in quantum dots, most notably the discrete time crystal (DTC). 
        This is significant given that spin qubits have fallen behind other qubit architectures in terms of size and control. 
        However, silicon spin qubits are especially well suited to this task, as the charge noise that usually foils gate operations can now be leveraged as an asset in this time crystal realization. 
        We illustrate differences between prethermal phenomena and true time-crystalline spatiotemporal order. 
        We demonstrate that even for a spin chain of four qubits, rich regime structures can be established by observing signatures of the discrete time crystal and the Floquet symmetry-protected topological regime (FSPT), both distinct from the thermal regime. 
        We also analyze the persistence of these signatures at longer chain lengths, showing that the DTC lifetime grows exponentially with the system length, and that these signatures may even be detectable for chains as small as three qubits. 
        We also discuss the effects of longer pulse durations and the effectiveness of pulse sequences for converting the exchange interaction to an Ising model.
        Our theoretical predictions are well-suited for immediate experimental implementations using currently existing quantum dot spin qubit systems.
    \end{abstract}
    \maketitle

    \section{Introduction}
    \label{sec:Intro}

        Spontaneous symmetry breaking occurs when the gro\-und state of a system does not obey the symmetries of its corresponding Hamiltonian. 
        For almost every possible symmetry of a Hamiltonian, there are known instances of spontaneous symmetry breaking.
        One notable exception to this rule has been time-translation symmetry.
        Such a system that spontaneously breaks time-translation symmetry has been dubbed a ``time crystal,'' analogous to ordinary crystals which spontaneously break spatial translation symmetry. 
        Since we live in a Lorentz invariant space-time, it seems reasonable that translational symmetry breaking be general to both space and time.
        The ideas behind time crystals have repeatedly captured the interest of physicists since the initial investigations of perpetual motion, and have recently attracted renewed attention, with the proposal of several concrete models which claim to display generic time-translation symmetry breaking (TTSB) \cite{Wilczek2012,Li2012}. 
        However, the discovery of various no-go theorems showed that such a generic spontaneous-symmetry-breaking time crystal is impossible to realize in nature \cite{Bruno2013,Watanabe2015}.
        A more narrow definition of time crystals is based on Floquet theory featuring time-periodic Hamiltonians $H(t + T) = H(t)$. These are called ``discrete time crystals'' (DTCs) \cite{Khemani2016,Else2016,Yao2017,Ippoliti2021}.
        DTCs do not spontaneously break the invariance of \emph{continuous} time translations, only \emph{discrete} translations. 
        Thus, the Hamiltonian for a discrete time crystal, known as a Floquet Hamiltonian, has period $T$, but a DTC ground state locks into dynamics that are periodic with a period that is an integer multiple of the Floquet period $nT$, even when there is appreciable noise in the Floquet drive.
        The most commonly investigated form of DTC involves period doubling ($n=2$).
        
        Discrete time crystals utilize many-body localization (MBL) to remain stable in the thermodynamic limit \cite{Huse2015,Imbrie2016,Ponte2015}. This would make the discrete time crystal a non-equilibrium, dynamical phase of matter. 
        Rather than evaluating a spatial order parameter only after the system reaches thermal equilibrium, the order parameter for this phase would be the temporal autocorrelators of each individual constituent of the system, the system-wide order becoming \emph{spatiotemporal}.
        A DTC phase would never reach thermal equilibrium, and so it does not have a ``ground state'' in the traditional sense.
        Instead, the eigenstates of the Floquet unitary would be approximate linear combinations of $n$ energy eigenstates of the static Hamiltonian \cite{Else2016}. 
        For example, in the case of a many-body electron spin system, the Floquet eigenstates would be an approximate linear combination of $n$ different spin eigenstates (an eigenstate of $\prod \sigma^z_i$) and each spin eigenstate would subsequently act as a Floquet ``ground state,'' oscillating between $n-1$ other spin eigenstates with each application of $U_F$, the Floquet unitary. 
        Normally, a Floquet system would absorb energy from this drive and eventually thermalize, but for DTCs, this thermalization is avoided through many-body localization, and oscillations continue for the DTC lifetime, which diverges exponentially with system size $L$. 
        For a truly many-body system, this DTC lifetime well exceeds the lifetime of the observable universe. 
        Even chains of intermediate length have infinite lifetime for all practical purposes.
        Obviously, many-body systems that retain memory of their initial conditions for infinite times have great applications as a potential memory of a quantum computer.
        We will not discuss in this paper the important fundamental question regarding the existence or not of a true thermodynamic nonequlibrium DTC phase of matter (which at the minimum necessitates the highly debatable issue of the existence or not of true many body localization)-- our goal is to establish theoretically the possibility of the realization of an \emph{effective} DTC phase in small arrays of quantum dot based spin qubits for long times of experimental relevance.
         
        The emergence of a time scale that is not commensurate with the Hamiltonian of the system is not a new idea (or a quantum idea). 
        Everyday clocks are based on a time scale emergence. 
        Period doubling specifically is a common phenomenon that has been observed many times throughout the history of physics \cite{Faraday1831}.
        DTCs are unique in two ways, however. 
        First, they do not absorb energy from the environment or from the driving force.
        Hence, these ``clocks'' need no winding.
        Second, there is a finite region of Floquet parameter space which triggers the \emph{same} subharmonic response in the system.
        For example, a single electron spin in an initial state $|\!\!\uparrow\rangle$ could undergo a periodic time evolution unitary $U = \exp(\frac{-i\pi}{2}\sigma_x)$. 
        Such a unitary is perfect $\pi$-pulse and is applied every period $T$.
        The spin would exhibit a subharmonic response, with the spin periodically returning to the initial state with doubled period $2T$.
        However, consider a slight error in the $\pi$-pulse, so that the periodic unitary $U' = \exp(\frac{-i9\pi}{20}\sigma_x)$. 
        The subharmonic response would suddenly be very different, and the system would only return to its initial state after $20T$. However, for a time crystal, the periodicity of the system remains $2T$, despite the perturbation. 
        These nontrivial attributes of DTC apply to small systems and often for very long times (essentially infinite times for laboratory systems, which all have some dissipation to the environment limiting the definition of `infinite' time).  

        A note on terminology is necessary. 
        Since all systems examined in this paper are of small size ($L=4$ for our main simulations), it is shortsighted to refer to any of the dynamics reviewed in this paper as a definite signature of a thermodynamic \emph{phase}. 
        This distinction is especially important now that the idea of MBL itself, upon which the theoretical underpinnings of discrete time crystals rely, is currently being more carefully scrutinized, and some numerical investigations have failed to reproduce the results predicted by MBL \cite{Lazarides2015,Abanin2021}. It seems that at the very least, the region of phase space corresponding to many-body localization is in general much smaller than previously thought \cite{Morningstar2021,Sels2021,Tu2022}. 
        With this in mind, we classify a finite system that behaves as a DTC as being in a DTC or TTSB \emph{regime}, with its behavior in the thermodynamic limit unknown and ambiguous, although in App.~\ref{app:scaling} we find encouraging evidence that the DTC behavior scales appropriately for a thermodynamic phase, diverging exponentially with system size, though such extrapolations are always suspect without a proper analytical theory.

        We define a DTC or TTSB regime as one in which the system exhibits spatiotemporal order 
        (1) throughout the entire system (not just on the edges),
        (2) throughout a finite region of parameter space (not just for a small set of ``cherry-picked'' initial states), and
        (3) that is robust to perturbations, locking into the period doubling dynamics even when the spin-flipping pulses deviate from the ideal limit of $U = \exp(\frac{-i\pi}{2} \sigma_x)$.
        Obviously, in order for TTSB states to have an even somewhat plausible claim to extending to a thermodynamic phase, all of these requirements must be true. 
        We identify the system as being a Floquet symmetry-protected topological (FSPT) regime \cite{Khemani2016,Moessner2017,Harper2020,Zhang2022} when (2) and (3) are true, but not (1). The FSPT phase is a novel system in its own right, but is not the primary focus of this paper.

        Silicon quantum dot (QD) spin qubits are excellent candidates for universal quantum computation \cite{Loss1998}. 
        This is due to their long spin coherence time, fast gate times, and compatibility with existing advanced lithographic techniques for silicon-based classical processors, which provides certain advantages with regards to scalability. 
        However, it has been relatively difficult for researchers to develop silicon qubit systems with high gate fidelities over the years, compared with the rapid success of superconducting and trapped ion quantum computing architectures, although we have seen much recent progress in this regard \cite{Mills2021,Noiri2022,Philips2022}.
        Silicon QD qubits, however, have certain advantages compared to superconducting and trapped ion architectures in scalability. 
        This is important, as many of the disruptive applications of quantum computation, such as integer factorization, require thousands of logical qubits, which is equivalent to performing quantum error correction with about one billion physical qubits \cite{Fowler2012} (the qubits referenced throughout this paper are obviously physical spin qubits fabricated in quantum dots).
        Obviously for architectures other than silicon QD qubits, the technology required to fabricate such a large system is a major obstacle in itself, whereas the technology for a silicon quantum computer's scaling already exists to a certain degree \cite{Borsoi2022}.
        The problem with Si qubits, however, is  that very few qubit systems have been realized in the laboratory which can perform any actual quantum tasks of intrinsic interest.

        The current period in quantum computing has been dubbed the Noisy and Intermediate Scale Quantum (NISQ) device era, as the quantum devices that researchers are currently developing are too small and too noisy to solve important real world problems. 
        As stepping stones to universal quantum simulation and computation, many have begun pursuing the proverbial ``lower hanging fruits'' in analog quantum simulation in order to demonstrate a proof of concept. 
        Although these demonstrations have no immediate application (and often can actually be reproduced using large scale  digital simulations on high performance `classical' computing clusters), they nonetheless represent uncharted waters of science and technology, and should not be treated as anything less than that. 
        These types of demonstrations which are possible on intermediate-scale, noisy quantum devices represent important landmarks on the journey of building a universal quantum computer.
        Quantum simulations of a DTC have been performed in superconducting \cite{Mi2022}, trapped ion \cite{Zhang2017}, NMR \cite{Rovny2018}, and NV center \cite{Choi2017} qubits. 
        Achieving a similar demonstration of this landmark on a semiconducting spin qubit system would further solidify the promise of semiconductor spin qubits as a platform for future large-scale quantum computation.

        Silicon spin qubits are particularly well suited for simulating a discrete time crystal. 
        In order to simulate a DTC, an experimentalist needs only reasonably high-fidelity single-qubit gates and reliable initialization and readout. 
        Noise-free two-qubit gates are not necessary. 
        This is excellent news for silicon, since it has been shown \cite{Mills2021, Noiri2022,Philips2022} that it is possible to achieve high-fidelity for both single-qubit gates and initialization and readout.
        In fact, the issue that most commonly plagues silicon qubit systems (charge noise), is precisely the key ingredient to simulating a DTC (disorder that is invariant under an Ising symmetry $\hat{P} = \prod_i \sigma_i^x$, or ``Ising-even'')\cite{Ippoliti2021}.

        The rest of the paper is organized as follows. 
        We first introduce the mathematical model of our Floquet drive in Section \ref{sec:Model}. 
        We then present the results of our calculations in Section \ref{sec:Calculations}, illustrating several DTC ``red herrings'' and calculating ``regime diagrams'' for the DTC regime in silicon spin qubits. 
        We then summarize our findings in Section \ref{sec:Conc} and discuss their implications. 

   \section{Model}
   \label{sec:Model}

        Silicon QD spin qubits interact through the exchange interaction $H_\text{exch} = \sum_n J_n \boldsymbol{\sigma}_n \cdot \boldsymbol{\sigma}_{n+1}$, where $\boldsymbol{\sigma}_n$ is the Pauli spin vector for the $n^\text{th}$ qubit.
        However, in order to stabilize a time crystal, Ising interactions are necessary \cite{Barnes2019,Ippoliti2021}. 
        Therefore, we use a series of ``Heisenberg-to-Ising'' pulses (H2I) \cite{Barnes2019} to transform the exchange interaction to an Ising interaction. 
        We elaborate on this process in App.~\ref{app:H2I_pulses}. 
        
        Throughout this work, we use the following Floquet unitary, or time evolution operator, 
        \begin{equation*}
            U(t_0 + T; t_0) = U(T; 0) = U_F
        \end{equation*}
        which is composed of a binary drive
        \begin{equation}
            \label{eq:unitaries}
            U_F = U_2 U_1 = \exp({-i H_2} t_2)\exp(-i H_1 t_1),
        \end{equation}
        where
        \begin{align}
            H_1 &= \sum_n^L (1 - \varepsilon)  \frac{\pi}{2} \sigma^x_n \\ 
            H_2 &= \sum_n^{L-1} J_n \sigma^z_n \sigma^z_{n+1} + \sum_n^L h_n \sigma^z_n. 
        \end{align}
        By definition, the Floquet Hamiltonian is periodic over a Floquet period $T = t_1 + t_2$, so that $H(t + T) = H(t)$ and $U(t + T) = U_F U(t)$. 
        We choose $t_1=1$ so that 
        \begin{equation} 
            U_1 = \prod_n \exp\Big[-\frac{i \pi}{2} (1-\varepsilon)\sigma^x_n \Big]
        \end{equation}
        corresponds to an ideal spin flip, or $\pi$-pulse, of all qubits when $\varepsilon=0$. 
        When $\varepsilon$ is nonzero, this corresponds to error in the single qubit X-gate. This error can be intentional (to demonstrate stability to perturbations),  systematic (roughly corresponding to the single-qubit gate infidelity $1 - F \approx \varepsilon$, where $F$ is the single-qubit gate fidelity), or a combination of the two. 
        Choosing different values of $t_2$, while still maintaining that $U_1$ correspond to a $\pi$-pulse in the error-free case, can have appreciable effects on the outcome of the simulation. 
        Generally speaking, the longer $t_2$, the less disorder is required to stabilize TTSB. 
        We explore the effects of different pulse times $t_2$ in App.~\ref{app:H2I_pulses}.
        For all calculations in this work, unless otherwise noted, we choose $t_1=t_2=1$ and therefore $T=t_1+t_2=2$. As denoted in each relevant figure, the time is expressed in terms of Floquet period $T$, which is the natural time scale.

        In order to stabilize the DTC phase, it has been long recognized that disorder is a critical ingredient. 
        We implement two different forms of disorder, onsite disorder through magnetic noise and disorder in the Ising interaction through charge noise. 
        We found that the exact shape of disorder distribution was not important to the fidelity (see App.~\ref{app:noise}), so for simplicity both forms of disorder are drawn from uniform distributions so that $h_n \in [h_0 - \sigma_h, h_0 + \sigma_h]$ and  $J_n \in [\text{max}(J_0 - \sigma_J, 0) , J_0 + \sigma_J]$. 
        Ising interaction strengths are truncated to be nonnegative since the Ising interaction for silicon qubits is based on the exchange interaction.

        We define the $k^\text{th}$ autocorrelator of a spin chain after time $t$ as 
        \begin{equation}
            Z_k(t) = \min_n \Biggl(\;\Bigg|\Big\langle \sigma^z_k(0)\Big\rangle \Big\langle \sigma^z_k (t = 2nT) \Big\rangle\Bigg|\;\Biggr),
        \end{equation}
        where $n$ is an integer, and the brackets indicate the expectation value at a specific time in the dynamics. 
        This metric captures the maximum deviation of a spin from its DTC-predicted orientation, with $Z_k(t) = 0$ corresponding a to complete loss of information in the $k^\text{th}$ spin after time $t$.
        We define the lifetime of the $k^\text{th}$ spin as the time $t$ when $Z_k(t) < 0.1$.

        In the absence of disorder, the system will not show any period doubling for nonzero $\epsilon$. 
        When $\varepsilon \neq 0$, only a ``spin-echo'' effect is apparent, with the autocorrelator being periodic in $2T/\varepsilon$.
        However, this is a phenomenon distinct from DTC dynamics, since the period \emph{changes} as a function of $\varepsilon$, as opposed to locking onto a period of $2T$ for small $\varepsilon$.

    \section{Calculations}
    \label{sec:Calculations}

        \begin{figure*}
            \includegraphics*[scale=0.30]{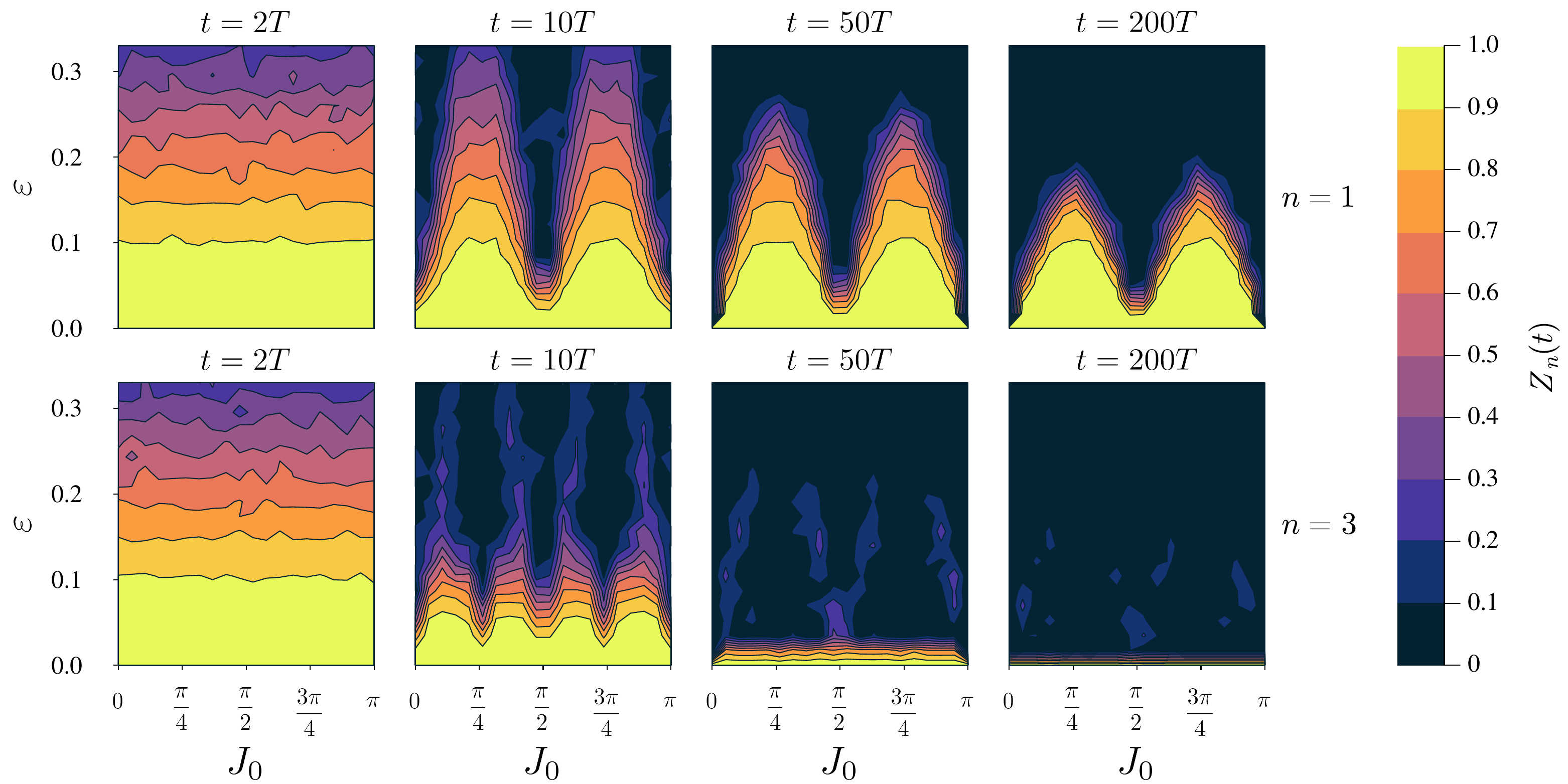}
            \caption{Average autocorrelators for the initial state $\ket{\psi_0} = \ket{1000}$  over a region of Ising couplings $J_0$ and $\pi$-pulse errors $\varepsilon$ in the absence of charge noise. 
            $h_0 = 2.0\times10^4$. 
            $\sigma_h=50.0$. 
            $\sigma_J =0.0$.  
            Averaged over 100 realizations of disorder. 
            \textbf{Top row:} The average autocorrelator of the first qubit $Z_1(t)$. 
            Two regions of high autocorrelation persist after 200 Floquet periods. 
            \textbf{Bottom row: } The average autocorrelator of the third qubit $Z_3(t)$. 
            Here we see that any high autocorrelation regions quickly vanish when the autocorrelator belongs to a bulk spin, rather than an edge spin.}
            \label{fig:noNoiseDynamics}
        \end{figure*}

        We apply the Floquet unitary repeatedly on a chain of four qubits, which is a system size that is currently accessible to state of the art devices \cite{Sigillito2019,Lawrie2020}. 
        After a predetermined number of Floquet periods, we calculate the autocorrelator $Z_n(t)$ for the $n^\text{th}$ qubit. 
        We repeat this calculation for various values of $\varepsilon$ and $J_0$, highlighting regions where spatiotemporal order is present. 
        In order to differentiate the DTC regime from other dynamics, it is necessary to confirm that spatiotemporal order persists for a variety of choices for both initial state and $n$.

        Throughout this work, we examine ``regime diagrams,'' in which we perform a parameter sweep over the $(\varepsilon, J_0)$ plane at a certain time $t$, and distinguish regimes with DTC behavior from those that show no such response. 
        Dynamics with high autocorrelation will show a region with $Z_n(t) \approx 1.0$, which will correspond to a yellow region on our regime diagrams.
        The DTC regime is one in which these yellow regions of autocorrelation are persistent throughout the chain, for long times, and in the presence of the perturbation $\varepsilon$.
    
\begin{figure*}
            \includegraphics*[scale=0.30]{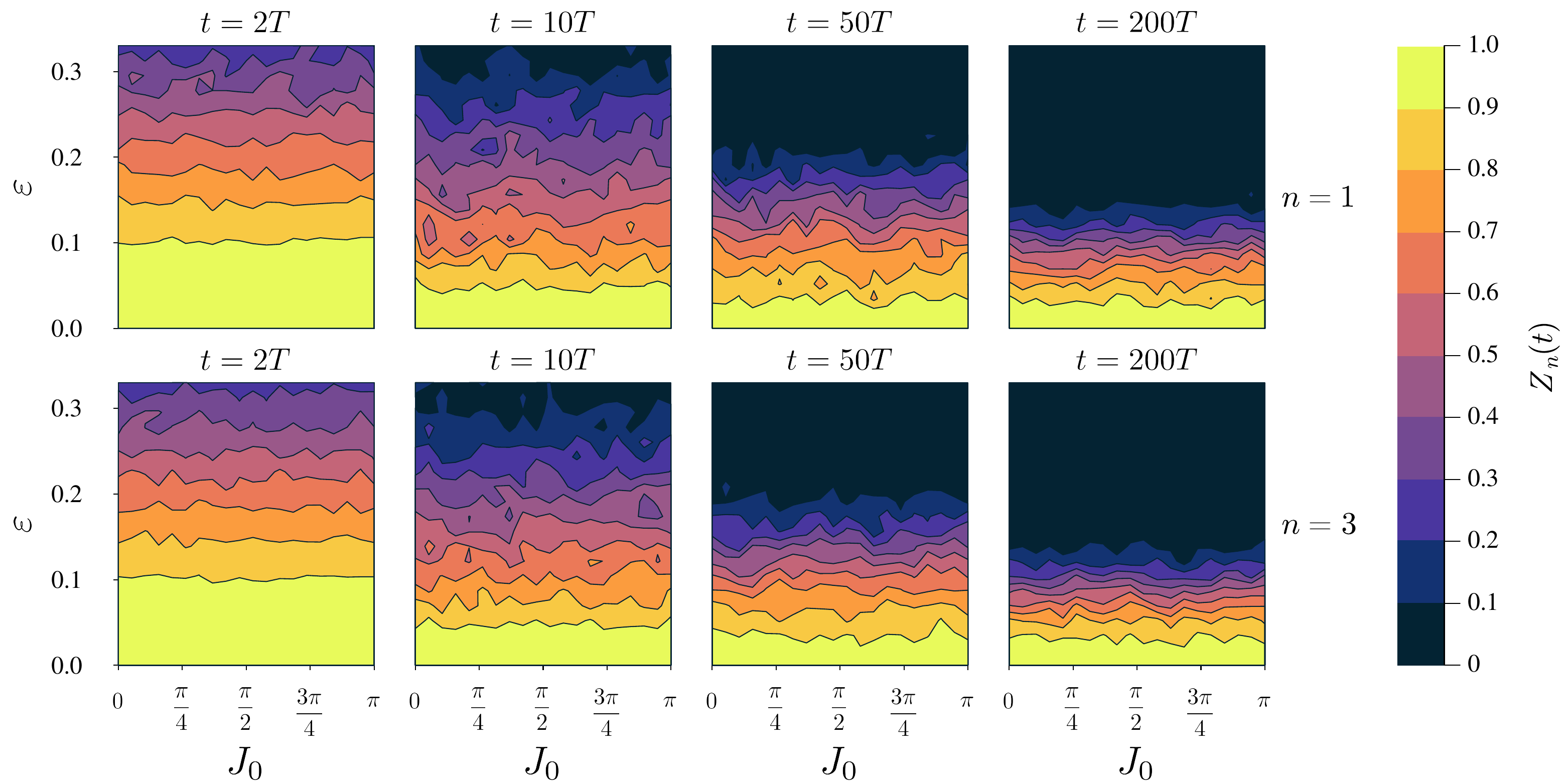}
            \caption{Average autocorrelators for the initial state $\ket{\psi_0} = \ket{1000}$  over a region of Ising couplings $J_0$ and $\pi$-pulse errors $\varepsilon$ in the presence of charge noise.
            $h_0 = 2.0\times10^4$. 
            $\sigma_h=50.0$. 
            $\sigma_J = 3.0$.  
            Averaged over 100 realizations of disorder. 
            \textbf{Top row:} The average autocorrelator of the first qubit $Z_1(t)$. 
            For sufficiently small $\varepsilon$, we see high autocorrelation for any value of $J_0$. 
            \textbf{Bottom row: } The average autocorrelator of the third qubit $Z_3(t)$. 
            Because of charge noise, the DTC regime is apparent for either spin observed. 
            The autocorrelators for both bulk and edge spins are about the same.}
            \label{fig:wNoiseDynamics}
        \end{figure*}

        We examine the dynamics of the spin chain without charge noise in Fig.~\ref{fig:noNoiseDynamics}. 
        Here we see the first of many ways researchers could potentially be deceived into thinking they have stabilized a time-crystalline state. 
        This is one example of a feature that we observed in every single system we examined: an edge spin ($n=1$ or $n=L$) always shows a dramatically longer lifetime than any spin in the bulk, provided $\varepsilon$ is not so large as to destroy all autocorrelation.
        This discrepancy between edge and bulk qubit lifetimes is indicative of a Floquet symmetry-protected topological regime, rather than a DTC regime.
        In the absence of sufficient charge noise, it is not uncommon to have robust spatiotemporal order in the edge spins while the bulk has completely thermalized. 
        We perform the same calculations with charge noise in Fig.~\ref{fig:wNoiseDynamics}. 
        Here we see the persistence of a finite region of parameter space that exhibits time-translation symmetry breaking, irrespective of which autocorrelator we examine. 
        In the absence of charge noise, the region of subharmonic response quickly vanishes.
        Once charge noise is included, the subharmonic response region shrinks until about $t = 50T$. 
        From then on, it remains approximately constant for long times ($t \approx 1000T$).

        Clearly Fig.~\ref{fig:noNoiseDynamics} does not represent a state in a DTC regime. 
        The absence of charge noise leads to the bulk thermalizing after a few tens of Floquet periods. 
        The importance of charge noise in establishing a DTC regime has been long known, but Figs.~\ref{fig:noNoiseDynamics}-\ref{fig:wNoiseDynamics} serve as a concrete demonstration of DTC dynamics for a small system size ($L=4$) and small times ($t<200T$). 
        These regimes can be explored using current semiconductor spin qubit technology, especially since only high-fidelity readout, initialization, and single-qubit gates are necessary.

\begin{figure*}
            \includegraphics*[scale=0.30]{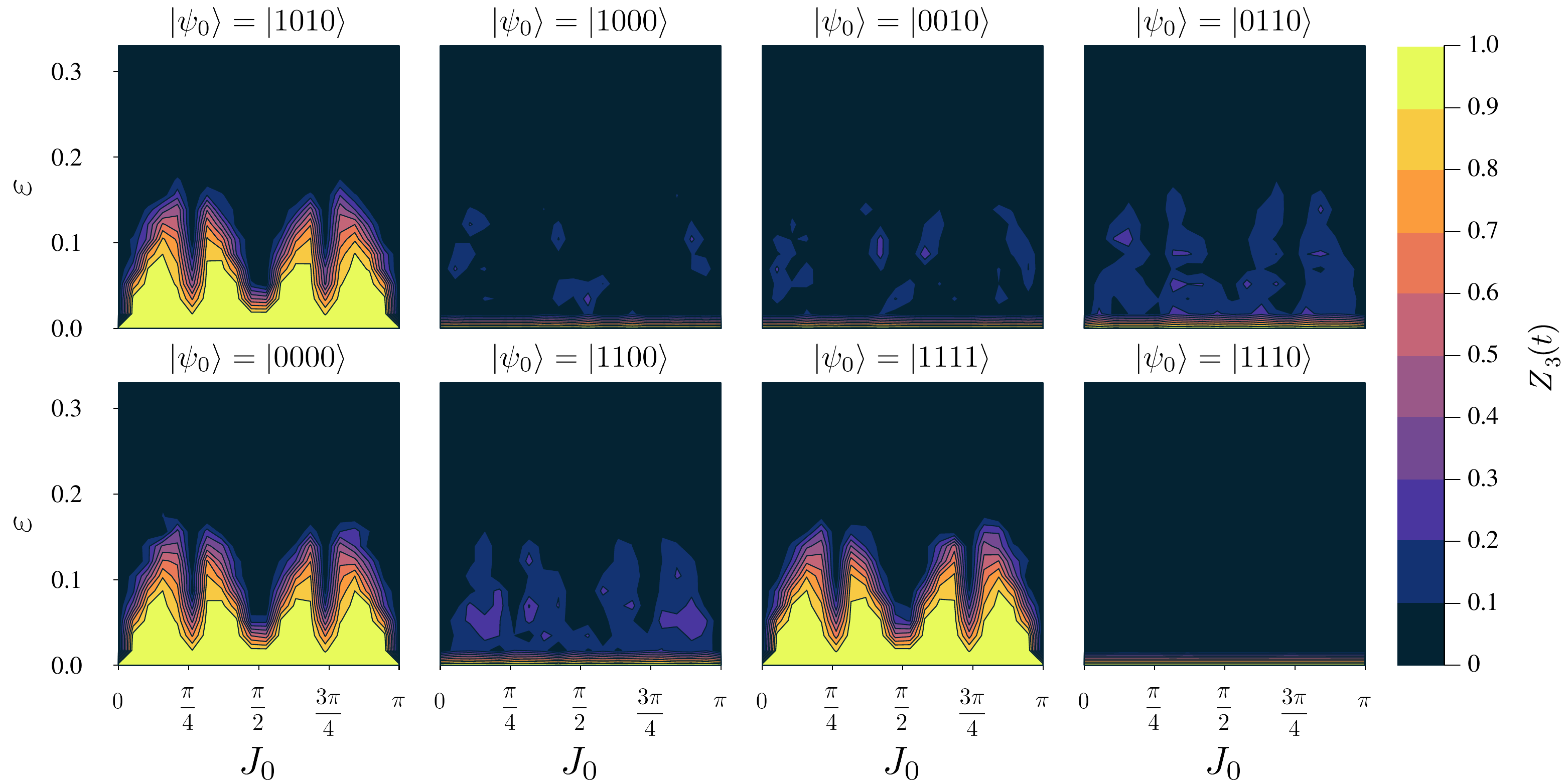}
            \caption{The autocorrelator $Z_3(t)$ for various initial states in the absence of charge noise. 
            $h_0 = 2.0\times10^4$.
            $t = 200T$. 
            $\sigma_h=50.0$. 
            $\sigma_J = 0.0$. 
            Averaged over 100 realizations of disorder.
            Ferromagnetic and antiferromagnetic states always exhibit regions of high autocorrelation for all spins, regardless of whether the system is in a DTC regime. 
            However, for a system to be in the DTC regime, it must exhibit high autocorrelation for any random initial state.}
            \label{fig:noNoiseStates}
        \end{figure*}

\begin{figure*}
            \includegraphics*[scale=0.30]{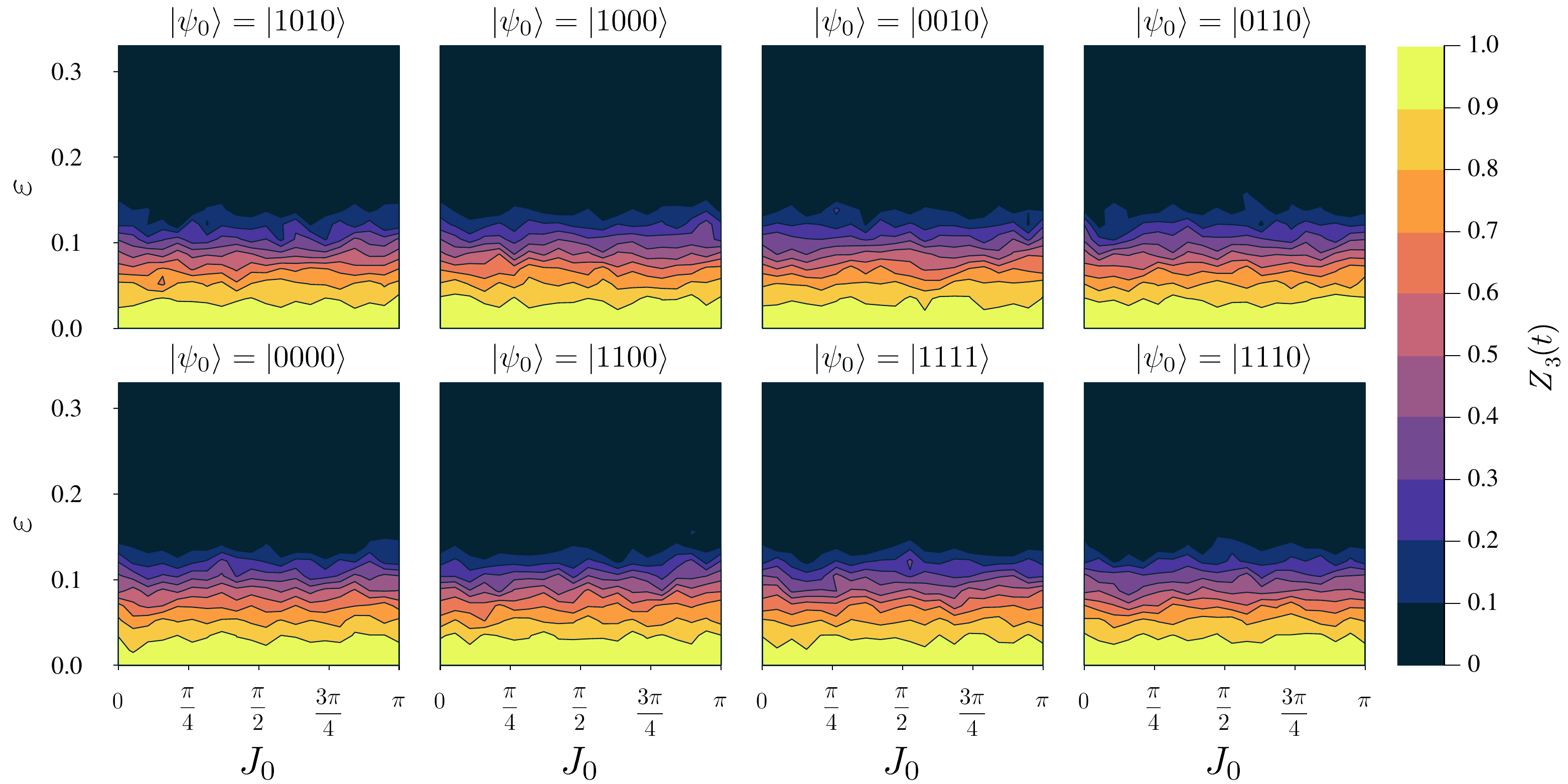}
            \caption{The autocorrelator $Z_3(t)$ for various initial states with the inclusion of charge noise. 
            $h_0 = 2.0\times10^4$. 
            $t = 200T$.
            $\sigma_h=50.0$. 
            $\sigma_J = 3.0$. 
            Averaged over 100 realizations of disorder.
            This system exhibits high autocorrelation for any random initial state, which is one of several effective diagnostics for identifying the DTC regime.}
            \label{fig:wNoiseStates}
        \end{figure*}

        We can see a similar stabilizing effect of charge noise in Figs.~\ref{fig:noNoiseStates} and \ref{fig:wNoiseStates}.
        We examine the third autocorrelator, since only the bulk spins contain useful information regarding whether or not the system is in a DTC regime.
        Without charge noise, the autocorrelator generally vanishes for any initial state that is not ferromagnetic or antiferromagnetic. 
        When charge noise is present, a region of high autocorrelation persists for small $\varepsilon$, with no apparent dependence on the value of $J_0$.
        For larger systems, if the $n^\text{th}$ autocorrelator is in a ferromagnetic or antiferromagnetic domain and far from a domain wall, it is also possible to see longer than expected lifetimes, even though the system is not in a DTC regime. 
        Fortunately for $L=4$, our calculations are not affected by the effects of large, polarized domains.
        Here we see another ``red herring'' that may lead researchers astray. 
        If an experimentalist chooses an initially polarized state, as is common, the system will always be found to show subharmonic response for certain choices of $\varepsilon$ and $J_0$. 
        A time crystal should not display qualitatively different dynamics based on its initial configuration.

        In order to properly diagnose whether a spin chain is in a DTC regime, it is imperative both to measure autocorrelators in the bulk and to examine a variety of random initial states. 
        Edge autocorrelators have long lifetimes for all initial states. 
        Ferromagnetic and antiferromagnetic initial states have long lifetimes for all autocorrelators. 
        But neither are indicative of TTSB.

        When subharmonic response persists for the bulk spins of systems that lack any kind of initial spatial order, the system is in a DTC regime. 
        These calculations serve as clear evidence that DTC dynamics can be captured on a small number of qubits.  
        
\begin{figure}
            \includegraphics[scale=0.245]{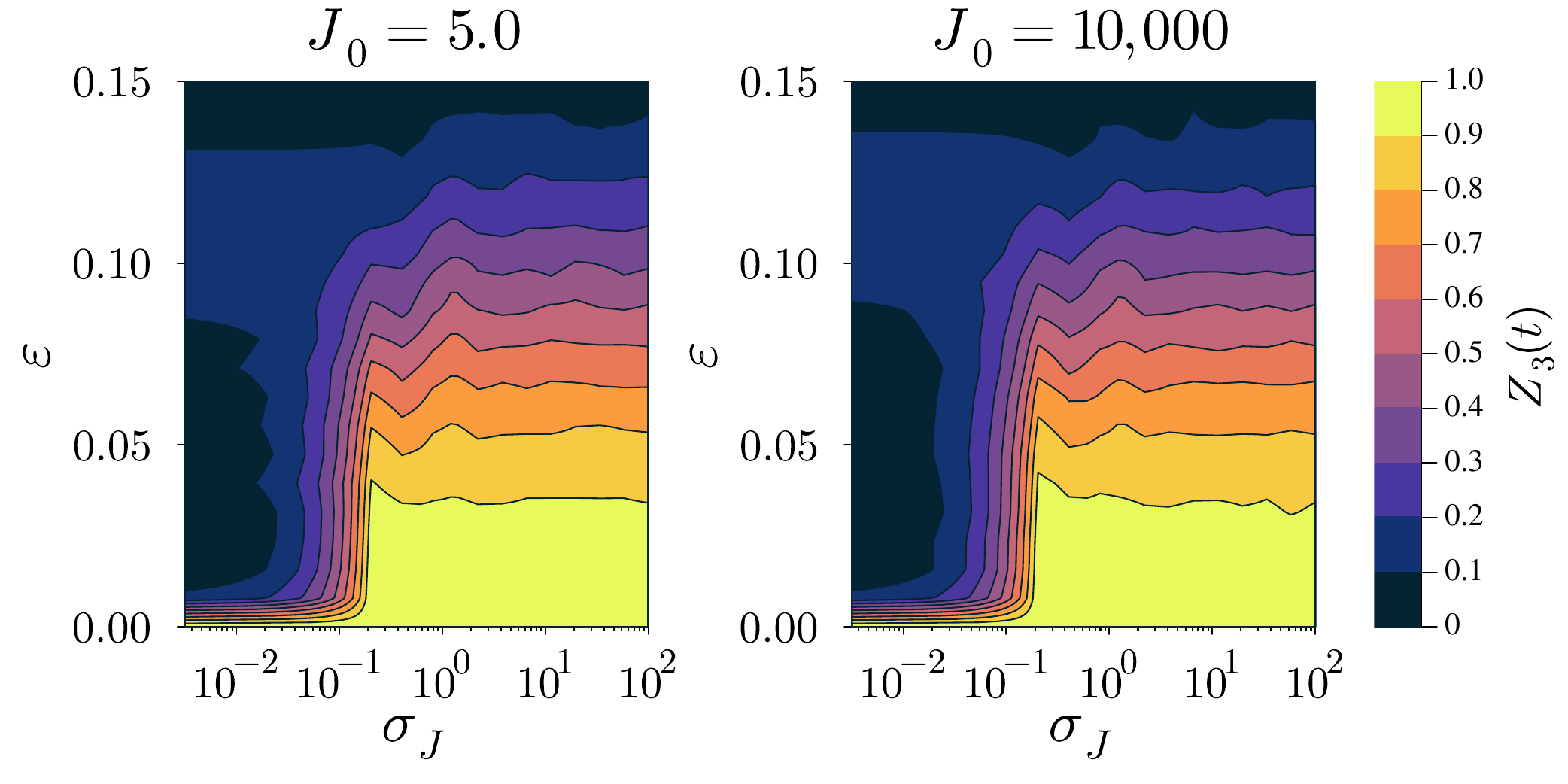}
            \caption{Regime diagrams for the DTC regime over a region of $(\varepsilon,\sigma_J)$ space. 
            $\ket{\psi_0} = \ket{1000}$, $t_2=1$, $h_0 = 2.0\times10^4$, and $\sigma_h =50.0$ for both plots. 
            A region of high autocorrelation clearly exists for small enough values of $\varepsilon$ when $\sigma_J t_2 \gtrsim 0.2$. 
            The autocorrelators are calculated after 200 Floquet periods and averaged over 2000 realizations of disorder. 
            The diagram on the left corresponds to $J_0 = 5.0$. 
            The diagram on the right, practically identical to the one on the left, corresponds to $J_0 = 10,\!000$.}
            \label{fig:sigJPhase}
        \end{figure}

\begin{figure}
            \includegraphics[scale=0.245]{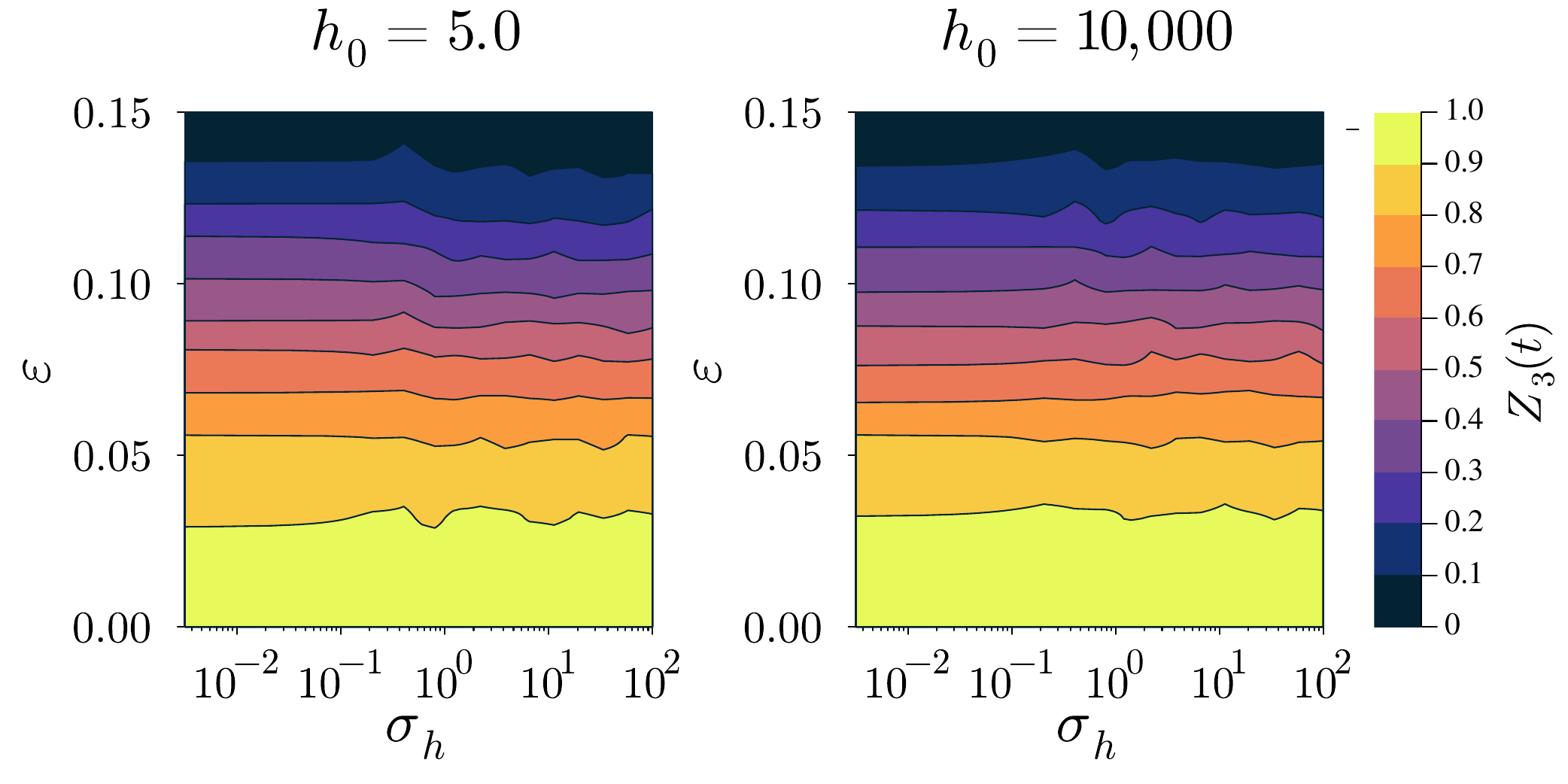}
            \caption{Regime diagrams for the DTC regime over a region of $(\varepsilon,\sigma_h)$ space.
            $\ket{\psi_0} = \ket{1000}$, $t_2=1$, $J_0 = 1.5$, and $\sigma_J = 3.0$ for both plots. 
            The autocorrelators are calculated after 200 Floquet periods and averaged over 2000 realizations of disorder. 
            The diagram on the left corresponds to $h_0 = 5.0$. 
            The diagram on the right, practically identical to the one on the left, corresponds to $h_0 = 10,\!000$.}
            \label{fig:sigHPhase}
        \end{figure}

        We examine the effects of charge noise $\sigma_J$ in a Fig.~\ref{fig:sigJPhase}. 
        A qualitative change is apparent at $\sigma_J \approx 0.2$. 
        The amount of charge noise necessary to stabilize the DTC regime appears to be independent of the base interaction strength $J_0$. 
        Sufficiently small $\sigma_J$ leads to the DTC regime completely vanishing, except for when $\varepsilon =0.0$, which is trivial, and not the signature of DTC dynamics.
        We examine the effects of magnetic noise $\sigma_h$ in a Fig.~\ref{fig:sigHPhase}. 
        The magnetic noise has little effect on the lifetime of the DTC regime because the disorder is Ising-odd, so the effects of the disorder are mostly echoed out after every two $\pi$-pulses. 
        The magnitude of $h_0$ has little to no effect on the regime of the system.

\begin{figure}
            \includegraphics*[scale=0.25]{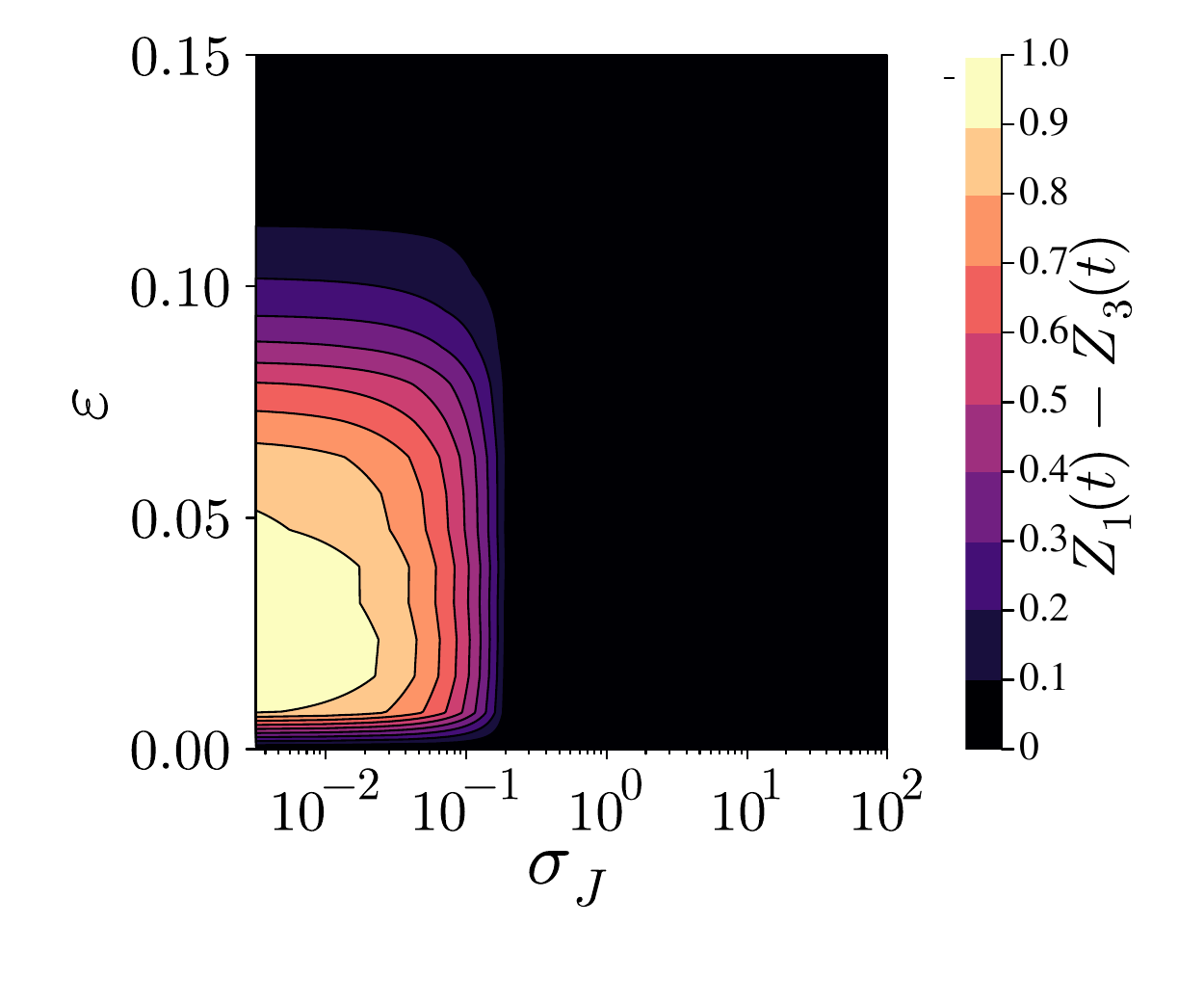}
            \caption{The difference between the first and third autocorrelator for an $L=4$ Ising spin chain. 
            The cream-colored region indicates the presence of edge modes where the bulk thermalizes ($Z_3(t) \approx 0$), but the edge spins retain their period doubling temporal order ($Z_1(t) \approx 1$). This is a signature of a Floquet symmetry-protected topological (FSPT) phase.}
            \label{fig:FSPT}
        \end{figure}

        The difference between an edge and bulk autocorrelator is shown in Fig.~\ref{fig:FSPT}. 
        An FSPT regime is clear for sufficiently small charge noise $\sigma_J$ and pulse error $\varepsilon$. 
        This agrees well with the findings in Fig.~\ref{fig:noNoiseDynamics}, and it is now clear that an FSPT regime survives beyond the limiting case where $\sigma_J=0$.
        For larger values of $\varepsilon$, the system completely thermalizes ($Z_n(t) \approx 0$  for all  $n$). 
        However, as we increase charge noise, the systems transitions from an FSPT regime to the DTC regime shown in Fig.~\ref{fig:sigJPhase}.

    \section{Summary and Conclusion}
    \label{sec:Conc}

    We examined the effect of a Floquet drive on autocorrelators of a spin chain of four qubits in an Ising model with nearest-neighbor, Ising-even disorder as well as onsite, magnetic disorder.
    We demonstrated that certain measurements are much more predictive of DTC order than others. 
    Whenever the initial state is spatially ordered, each autocorrelator in the chain exhibits spatiotemporal order for reasonable values of $\varepsilon$. 
    Edge autocorrelators are not able to distinguish between FSPT and DTC regimes for any initial state.
    Therefore, experimenters should always use local measurements in the bulk over a wide variety of initial states before claiming an observation of DTC order.
    We showed that significant charge noise is necessary in order to stabilize a DTC in semiconducting spin qubit systems.
    Insufficient charge noise can lead to an FSPT regime, but not a DTC regime.
    We explored several regime diagrams of such a system, varying the levels of $\pi$-pulse perturbation $\varepsilon$, charge noise $\sigma_J$, and magnetic noise $\sigma_h$.
    We found that the correct choices of the former two parameters are critical to stabilizing a DTC, but that magnetic noise was far less important.
    We simulated these phenomena by only considering a system size of $L=4$, which is well within the current spin qubit experimental capabilities. 

    
    Our scaling analysis extended the length of the qubit chains to $L=13$.
    Although this is far from the thermodynamic limit, the trend is encouraging, as the lifetime seems to diverge exponentially with the chain length.
    However, the possibility of demonstrating rich regime structure with only $L=4$ is an incredibly encouraging result for semiconducting spin qubits.
    Furthermore, simulating a DTC only requires reliable initialization and readout, and relatively high fidelity one-qubit gates (less than ideal one-qubit gates correspond to nonzero $\varepsilon$). 
    Noisy two-qubit gates should not inhibit the DTC realization, as Ising-even disorder is actually a key ingredient for system-wide subharmonic response. 

    Current silicon spin quantum devices are not capable of large-scale quantum computation due to their relatively small system sizes and prominent noise. 
    However, our work shows that they are capable of simulating complicated quantum dynamics which are not otherwise easily accessed experimentally.
    In the case of time crystals, these dynamics are surprisingly predicted to survive (possibly) in the thermodynamic limit, making them an out-of-equilibrium, dynamical phase of matter.
    Our work provides encouraging evidence that the stabilization of both FSPT and DTC regimes is currently accessible using modern QD spin qubit devices. 
    It also establishes that it is possible to explore prethermal dynamics in such devices for ``bad actor'' states which appear to be DTCs at first glance, but whose DTC order quickly deteriorates away from ``sweet spots'' in the choice of initial state or observable measured. 
    Distinguishing between these regimes is necessary in order to verify that any experimental dynamics are indicative of a discrete time crystal.
    Because these two regimes are distinguishable for both small system sizes and for short times, silicon quantum dot qubit systems are well positioned to host a physical realization of time crystals.

    \begin{acknowledgments}
        This work is supported by the Laboratory for Physical Sciences. 

    \end{acknowledgments}

    \appendix

    \section{Scaling Analysis}

    \label{app:scaling}

\begin{figure}
        \includegraphics*[scale=0.35]{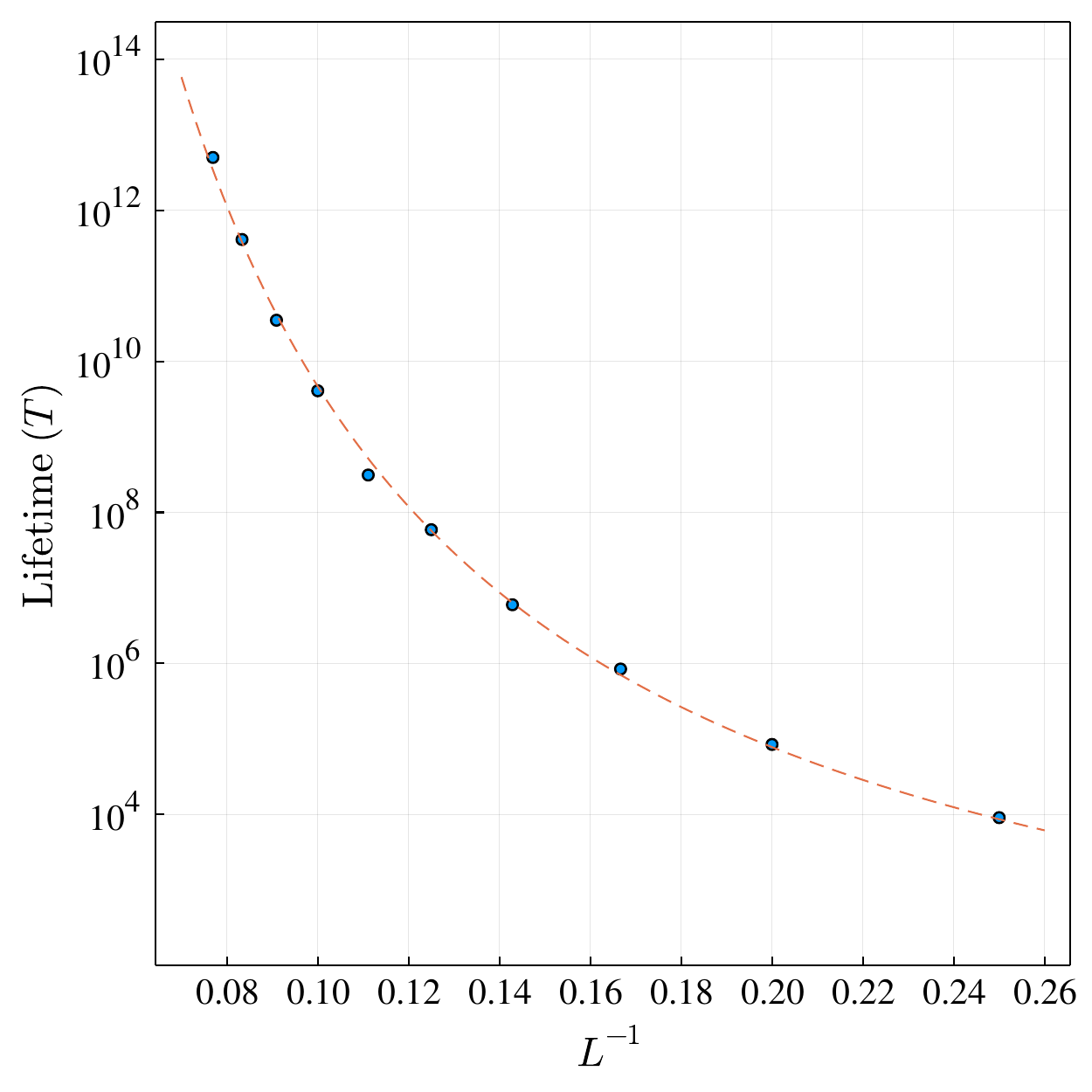}
        \caption{The average DTC lifetime for an Ising spin chain of length $L$. 
        The DTC lifetime is defined as the time at which the $n=L/2$ autocorrelator drops below 0.1 when $\varepsilon=0.10$. The dashed red line is the least-squares fit.
        }
        \label{fig:lifetimes}
    \end{figure}

    We performed scaling analysis for the average lifetime of a chain of length $L$. 
    After calculating $Z_{L/2}(t)$ over random initial states and many realizations of disorder, we found the average lifetime at which the autocorrelator dropped below 0.1. 
    These average lifetimes are plotted as a function of $1/L$ in Fig.~\ref{fig:lifetimes}. 
    We performed a logarithmic least-squares fit and found that the lifetimes approximately obeyed the following scaling behavior.
    \begin{equation*}
        \langle t_L \rangle = 1.3 \exp\Big(2.2 L\Big)
    \end{equation*}

    It's clear from Fig.~\ref{fig:lifetimes} that our simulations reproduce the theoretical prediction that the DTC lifetimes diverge exponentially with the system size.
    Of course, this says little regarding the behavior for truly many-body systems, whose number of constituents is much closer to Avogadro's number than to unity.
    However, it does show that for intermediate size systems, the DTC lifetime is infinite for all practical purposes. For a spin chain of $L=13$ that is completely isolated, the time-crystalline order is expected to survive for over 5 trillion Floquet periods before thermalizing. 

\begin{figure}
        \includegraphics[scale=0.26]{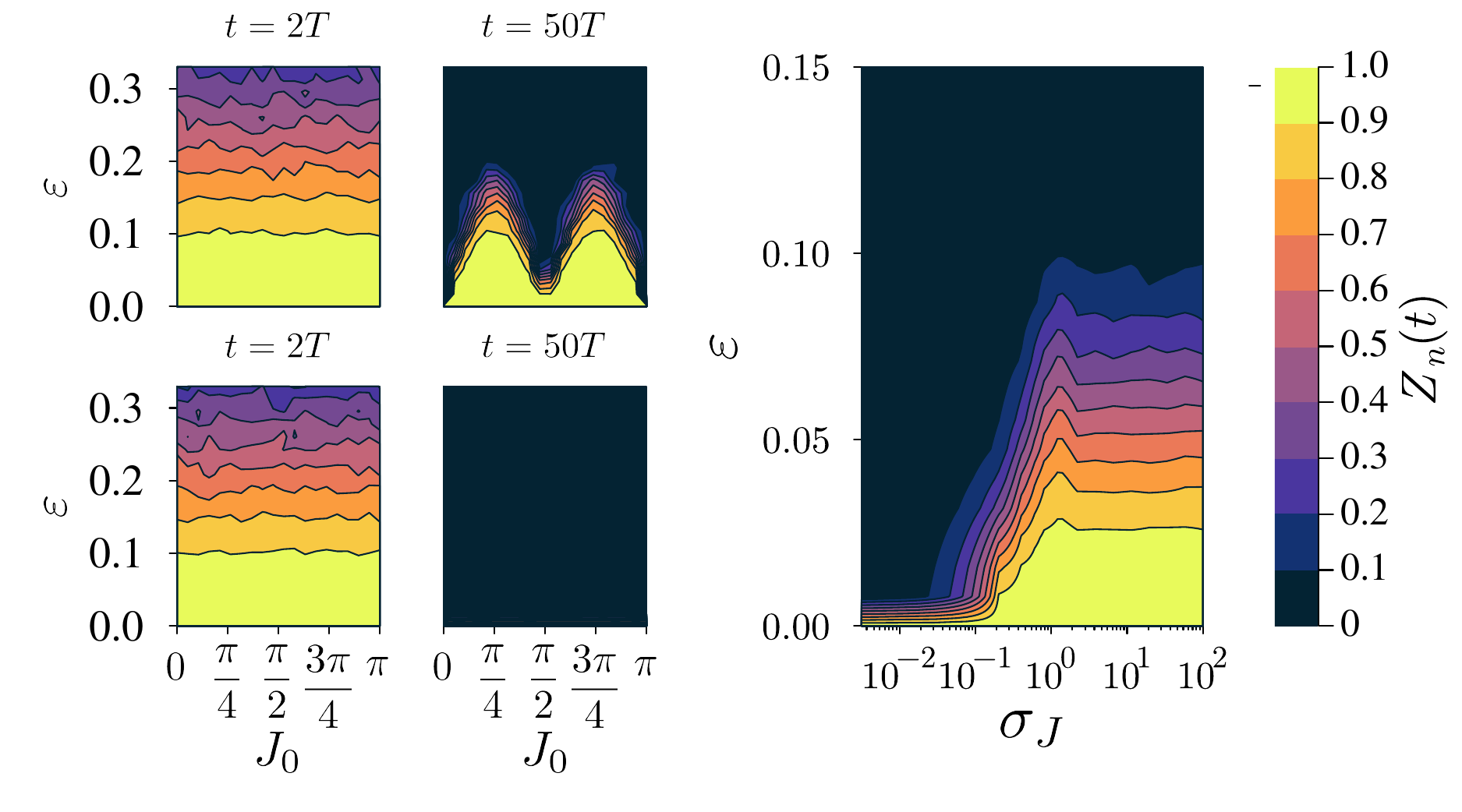}
        \caption{Various autocorrelators for $L=3$. 
        In each case, $\ket{\psi_0} = \ket{100}$, $\sigma_h =50.0$, and $h_0 = 2.0\times10^4$. 
        \textbf{Top left:} The first ($n=1$) autocorrelator after 2 and 50 Floquet periods, respectively. 
        $\sigma_J=0$. 
        \textbf{Bottom left:} The second ($n=2$) autocorrelator after 2 and 50 Floquet periods, respectively. 
        $\sigma_J=0$. 
        \textbf{Right:} A regime diagram of the DTC regime for $L=3$ ($n=2$). 
        Averaged over 2000 realizations of disorder, after 200 Floquet periods.}
        \label{fig:smaller_chains}
    \end{figure}

    It may even be possible to observe these dynamics in an even smaller spin chain. 
    We simulated an $L=3$ spin chain under an Ising model and show our results in Fig.~\ref{fig:smaller_chains}. 
    We use some of the same the diagnostics used earlier to identify FSPT and DTC regimes in $L=4$ chains, only this time with $L=3$. In the DTC regime shown in the regime diagram of Fig.~\ref{fig:smaller_chains}, we also confirmed that the subharmonic response is not sensitive to the choice of initial state.

    \section{Heisenberg-to-Ising pulses and pulse durations}
    \label{app:H2I_pulses}

    Silicon spin qubits interact via the Heisenberg exchange interaction. 
    Heisenberg-to-Ising (H2I) pulses are pulse sequences designed to transform the interactions of a  system from the Heisenberg exchange interaction to a simple Ising interaction \cite{Barnes2019}. 
    The unitary arising from Heisenberg interactions between quantum dots takes the following form 
    \begin{equation}
        U_H(t) = \exp\Bigg[ -i t \left(\sum_n^{L-1} J_n \boldsymbol{\sigma}_n \cdot \boldsymbol{\sigma}_{n+1} + \sum_n^L h_n \sigma_n^z \right) \Bigg].
    \end{equation}

    We then interleave $\pi$-pulses on every other spin in the chain throughout the duration of $t_2$ so that $U_2$, previously defined in Eq.~\ref{eq:unitaries}, with $n$ H2I pulses is instead given by

    \begin{multline}
        \tilde{U}_2 = \Bigg[ \exp \left(\frac{i \pi}{2} (\sigma_z^1 + \sigma_z^3) \right) U_H(t_2/n)\;\; \times \\
        \exp \left(\frac{-i \pi}{2} (\sigma_z^1 + \sigma_z^3) \right) U_H(t_2/n) \Bigg]^{n/2}.
    \end{multline}

\begin{figure}
        \includegraphics[scale=0.30]{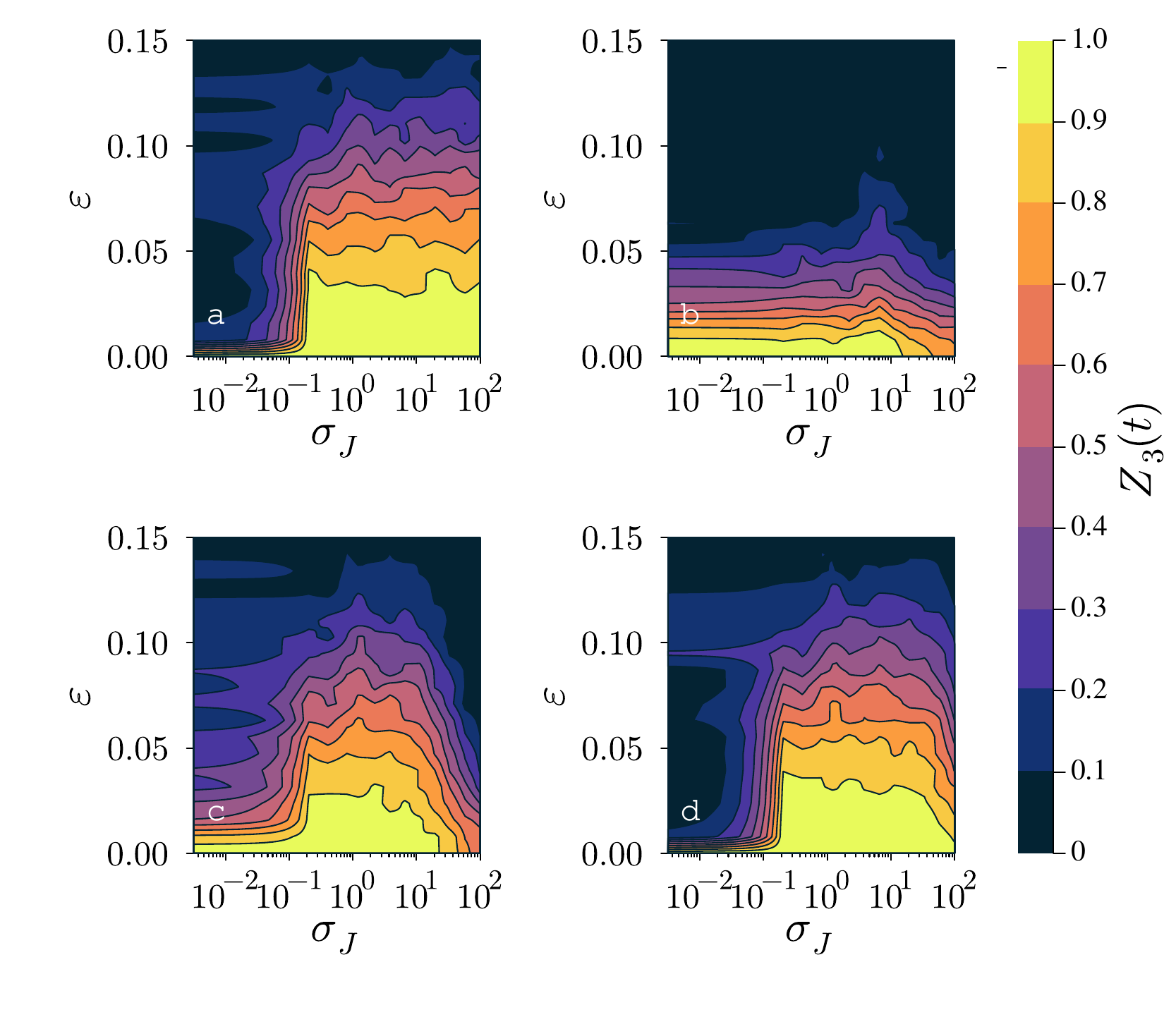}
        \caption{The effectiveness of Heisenberg-to-Ising (H2I) pulses. 
        The third autocorrelator $Z_3(t)$ after 200 Floquet periods, averaged over 200 realizations of disorder. 
        In all plots $\ket{\psi_0} = \ket{1000}$,  $J_0 =5.0$, $h_0 = 2.0\times10^4$, and $\sigma_h =50.0$. 
        \textbf{(a)} The regime diagram for an Ising model. 
        \textbf{(b-d)} The regime diagram for a Heisenberg model with \textbf{(b)} 8, \textbf{(c)} 64, and \textbf{(d)} 256 H2I pulses.}
        \label{fig:h2i_pulses}
    \end{figure}

    In an ideal case, where we have a pure Heisenberg model, then we would expect 2 H2I pulses to be sufficient to transform the system to an Ising model. 
    However, the presence of noise breaks this ideal case, and instead many pulses are needed to reproduce Ising-like results. 
    The effects of H2I pulses are shown in Fig.~\ref{fig:h2i_pulses}.

    Clearly, many H2I pulses are needed in order to well approximate the Ising case. 
    However, even when a large number of Heisenberg-to-Ising cannot be performed, a smaller number can  create a window of $\sigma_J$ which spans several orders of magnitude that can support a DTC regime, as seen in Fig.~\ref{fig:h2i_pulses}c-d.

\begin{figure}
        \includegraphics[scale=0.30]{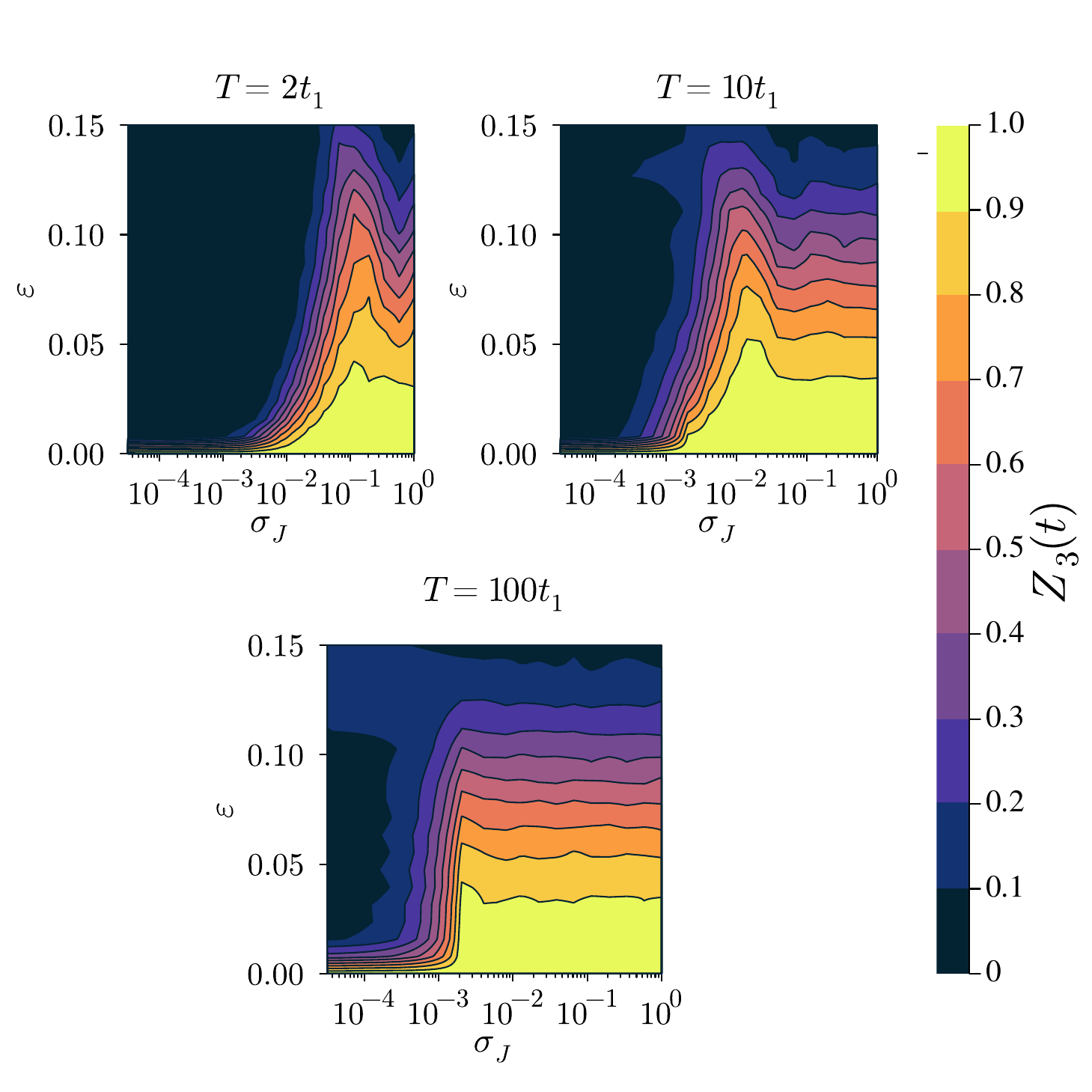}
        \caption{The effect of Floquet pulse duration. 
        Plotted are the DTC regime diagrams for different values of $T = t_1 + t_2$. 
        $t_2$ is varied, while $t_1$, the duration of the $\pi$-pulse, is held constant. 
        The regime diagrams are measured after 200 Floquet periods, and are averaged over 2000 realizations of disorder. 
        $\ket{\psi_0} = \ket{1000}$. 
        $J_0 =5.0$. 
        $h_0 = 2.0\times10^4$. 
        $\sigma_h = 50.0$.}
        \label{fig:pulse_durations}
    \end{figure}

    We investigated the effects of pulse timing, and found that the longer the ``stabilization'' period $t_2$, the less disorder is required in order to stabilize DTC order. 
    Our results are shown in Fig.~\ref{fig:pulse_durations}. In Fig.~\ref{fig:sigJPhase}, it was clear that the mean value of the Ising coupling had no effect on the boundaries of the DTC regime. 
    However, since the pulse duration scales the magnitude of the disorder, and not just the mean value, less disorder is required for larger values of $t_2$. According to our simulations, the transition from FSPT to DTC regimes occurs at $\sigma_J t_2 \approx 0.2$.
    Changing the minimum acceptable level of charge noise is especially helpful when the Ising interactions cannot be perfectly reproduced using H2I pulses. 
    In this case, one can extend $t_2$ in order to expand the window of acceptable levels of $\sigma_J$ in order to tune the DTC parameters to a specific quantum dot system.

    \section{Choice of exchange coupling disorder distribution}
    \label{app:noise}

    The Ising-even disorder (the critical ingredient of a discrete time crystal regime) originates from charge noise, which is prevalent in silicon quantum dots.
    Charge noise is a notoriously poorly understood topic from a theoretical standpoint, though we can still model charge noise phenomenologically.

    The charge noise power spectrum in silicon is approximately $1/f$ \cite{Reed2016, Connors2022}. 
    Accordingly, low-frequency dynamics dominate, and for short time scales, such as the lifetime of a four-qubit time crystal, it is reasonable to treat the noise as quasistatic, meaning that the exchange couplings can be treated as approximately constant. 
    This of course is not true for long time scales, such as for the DTC lifetime of a large number of qubits. 
    However, this would not affect the fact that DTC-like behavior can be detected and differentiated from the most common experimental ``red herrings'' using only four quantum dots in silicon.

    \begin{figure*}
        \includegraphics*[scale=0.30]{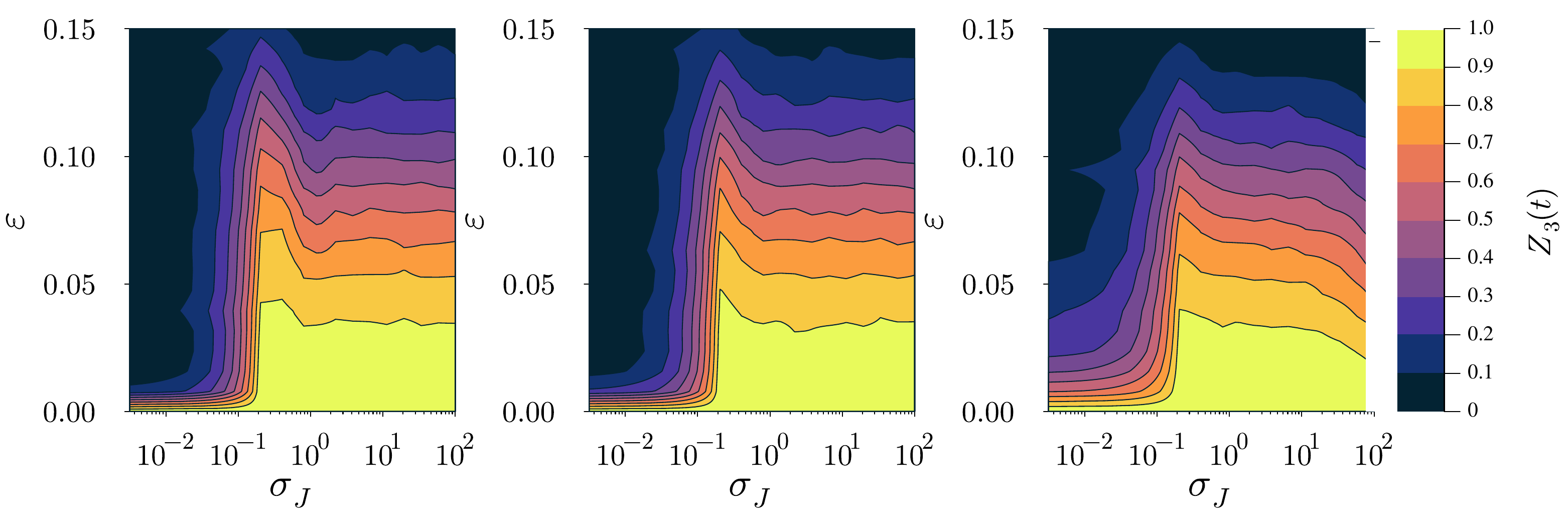}
        \caption{Autocorrelator regime plots showing the irrelevance of the exact shape in the exchange coupling probability distribution function.  
        \textbf{Left:} The exchange couplings $J_n$ are drawn from a uniform distribution of width $2 \sigma_J$. 
        \textbf{Middle:} The exchange couplings $J_n$ are drawn from a normal distribution with standard deviation $\sigma_J$.
        \textbf{Right:} The exchange couplings $J_n$ are drawn from a Cauchy-Lorentz distribution with full width at half maximum $2\sigma_J$.
        For all three plots, $J_0 = 200$, $|\psi_0\rangle = |1000\rangle$, $h_0 = 2.0\times10^4$, $\sigma_h = 50.0$, and the magnetic field probability distribution function remains $h_n \in [h_0 - \sigma_h, h_0 + \sigma_h]$.}
        \label{fig:noiseDists}
    \end{figure*}
    
    The next natural question to consider is whether the precise shape of the exchange coupling probability distribution affects our results.
    We repeated our calculations from Fig.~\ref{fig:sigJPhase}, but with different distributions of exchange coupling. We present these results in Fig.~\ref{fig:noiseDists}. 
    We considered three different probability distribution functions for the exchange couplings: uniform, normal, and Cauchy-Lorentz.
    The exact shape of the distribution function of $J_n$ had little qualitative effect on our main result: In order to detect the subharmonic response that is a true signature of time-crystalline behavior in such small systems, the disorder in the exchange couplings must be sufficiently large.

    \bibliography{references.bib}   

\begin{thebibliography}{35}%
\makeatletter
\providecommand \@ifxundefined [1]{%
 \@ifx{#1\undefined}
}%
\providecommand \@ifnum [1]{%
 \ifnum #1\expandafter \@firstoftwo
 \else \expandafter \@secondoftwo
 \fi
}%
\providecommand \@ifx [1]{%
 \ifx #1\expandafter \@firstoftwo
 \else \expandafter \@secondoftwo
 \fi
}%
\providecommand \natexlab [1]{#1}%
\providecommand \enquote  [1]{``#1''}%
\providecommand \bibnamefont  [1]{#1}%
\providecommand \bibfnamefont [1]{#1}%
\providecommand \citenamefont [1]{#1}%
\providecommand \href@noop [0]{\@secondoftwo}%
\providecommand \href [0]{\begingroup \@sanitize@url \@href}%
\providecommand \@href[1]{\@@startlink{#1}\@@href}%
\providecommand \@@href[1]{\endgroup#1\@@endlink}%
\providecommand \@sanitize@url [0]{\catcode `\\12\catcode `\$12\catcode
  `\&12\catcode `\#12\catcode `\^12\catcode `\_12\catcode `\%12\relax}%
\providecommand \@@startlink[1]{}%
\providecommand \@@endlink[0]{}%
\providecommand \url  [0]{\begingroup\@sanitize@url \@url }%
\providecommand \@url [1]{\endgroup\@href {#1}{\urlprefix }}%
\providecommand \urlprefix  [0]{URL }%
\providecommand \Eprint [0]{\href }%
\providecommand \doibase [0]{https://doi.org/}%
\providecommand \selectlanguage [0]{\@gobble}%
\providecommand \bibinfo  [0]{\@secondoftwo}%
\providecommand \bibfield  [0]{\@secondoftwo}%
\providecommand \translation [1]{[#1]}%
\providecommand \BibitemOpen [0]{}%
\providecommand \bibitemStop [0]{}%
\providecommand \bibitemNoStop [0]{.\EOS\space}%
\providecommand \EOS [0]{\spacefactor3000\relax}%
\providecommand \BibitemShut  [1]{\csname bibitem#1\endcsname}%
\let\auto@bib@innerbib\@empty
\bibitem [{\citenamefont {Wilczek}(2012)}]{Wilczek2012}%
  \BibitemOpen
  \bibfield  {author} {\bibinfo {author} {\bibfnamefont {F.}~\bibnamefont
  {Wilczek}},\ }\href {https://doi.org/10.1103/PhysRevLett.109.160401}
  {\bibfield  {journal} {\bibinfo  {journal} {Phys. Rev. Lett.}\ }\textbf
  {\bibinfo {volume} {109}},\ \bibinfo {pages} {160401} (\bibinfo {year}
  {2012})}\BibitemShut {NoStop}%
\bibitem [{\citenamefont {Li}\ \emph {et~al.}(2012)\citenamefont {Li},
  \citenamefont {Gong}, \citenamefont {Yin}, \citenamefont {Quan},
  \citenamefont {Yin}, \citenamefont {Zhang}, \citenamefont {Duan},\ and\
  \citenamefont {Zhang}}]{Li2012}%
  \BibitemOpen
  \bibfield  {author} {\bibinfo {author} {\bibfnamefont {T.}~\bibnamefont
  {Li}}, \bibinfo {author} {\bibfnamefont {Z.-X.}\ \bibnamefont {Gong}},
  \bibinfo {author} {\bibfnamefont {Z.-Q.}\ \bibnamefont {Yin}}, \bibinfo
  {author} {\bibfnamefont {H.~T.}\ \bibnamefont {Quan}}, \bibinfo {author}
  {\bibfnamefont {X.}~\bibnamefont {Yin}}, \bibinfo {author} {\bibfnamefont
  {P.}~\bibnamefont {Zhang}}, \bibinfo {author} {\bibfnamefont {L.-M.}\
  \bibnamefont {Duan}},\ and\ \bibinfo {author} {\bibfnamefont
  {X.}~\bibnamefont {Zhang}},\ }\href
  {https://doi.org/10.1103/PhysRevLett.109.163001} {\bibfield  {journal}
  {\bibinfo  {journal} {Phys. Rev. Lett.}\ }\textbf {\bibinfo {volume} {109}},\
  \bibinfo {pages} {163001} (\bibinfo {year} {2012})}\BibitemShut {NoStop}%
\bibitem [{\citenamefont {Bruno}(2013)}]{Bruno2013}%
  \BibitemOpen
  \bibfield  {author} {\bibinfo {author} {\bibfnamefont {P.}~\bibnamefont
  {Bruno}},\ }\href {https://doi.org/10.1103/PhysRevLett.111.070402} {\bibfield
   {journal} {\bibinfo  {journal} {Phys. Rev. Lett.}\ }\textbf {\bibinfo
  {volume} {111}},\ \bibinfo {pages} {070402} (\bibinfo {year}
  {2013})}\BibitemShut {NoStop}%
\bibitem [{\citenamefont {Watanabe}\ and\ \citenamefont
  {Oshikawa}(2015)}]{Watanabe2015}%
  \BibitemOpen
  \bibfield  {author} {\bibinfo {author} {\bibfnamefont {H.}~\bibnamefont
  {Watanabe}}\ and\ \bibinfo {author} {\bibfnamefont {M.}~\bibnamefont
  {Oshikawa}},\ }\href {https://doi.org/10.1103/PhysRevLett.114.251603}
  {\bibfield  {journal} {\bibinfo  {journal} {Phys. Rev. Lett.}\ }\textbf
  {\bibinfo {volume} {114}},\ \bibinfo {pages} {251603} (\bibinfo {year}
  {2015})}\BibitemShut {NoStop}%
\bibitem [{\citenamefont {Khemani}\ \emph {et~al.}(2016)\citenamefont
  {Khemani}, \citenamefont {Lazarides}, \citenamefont {Moessner},\ and\
  \citenamefont {Sondhi}}]{Khemani2016}%
  \BibitemOpen
  \bibfield  {author} {\bibinfo {author} {\bibfnamefont {V.}~\bibnamefont
  {Khemani}}, \bibinfo {author} {\bibfnamefont {A.}~\bibnamefont {Lazarides}},
  \bibinfo {author} {\bibfnamefont {R.}~\bibnamefont {Moessner}},\ and\
  \bibinfo {author} {\bibfnamefont {S.~L.}\ \bibnamefont {Sondhi}},\ }\href
  {https://doi.org/10.1103/PhysRevLett.116.250401} {\bibfield  {journal}
  {\bibinfo  {journal} {Phys. Rev. Lett.}\ }\textbf {\bibinfo {volume} {116}},\
  \bibinfo {pages} {250401} (\bibinfo {year} {2016})}\BibitemShut {NoStop}%
\bibitem [{\citenamefont {Else}\ \emph {et~al.}(2016)\citenamefont {Else},
  \citenamefont {Bauer},\ and\ \citenamefont {Nayak}}]{Else2016}%
  \BibitemOpen
  \bibfield  {author} {\bibinfo {author} {\bibfnamefont {D.~V.}\ \bibnamefont
  {Else}}, \bibinfo {author} {\bibfnamefont {B.}~\bibnamefont {Bauer}},\ and\
  \bibinfo {author} {\bibfnamefont {C.}~\bibnamefont {Nayak}},\ }\href
  {https://doi.org/10.1103/PhysRevLett.117.090402} {\bibfield  {journal}
  {\bibinfo  {journal} {Phys. Rev. Lett.}\ }\textbf {\bibinfo {volume} {117}},\
  \bibinfo {pages} {090402} (\bibinfo {year} {2016})}\BibitemShut {NoStop}%
\bibitem [{\citenamefont {Yao}\ \emph {et~al.}(2017)\citenamefont {Yao},
  \citenamefont {Potter}, \citenamefont {Potirniche},\ and\ \citenamefont
  {Vishwanath}}]{Yao2017}%
  \BibitemOpen
  \bibfield  {author} {\bibinfo {author} {\bibfnamefont {N.~Y.}\ \bibnamefont
  {Yao}}, \bibinfo {author} {\bibfnamefont {A.~C.}\ \bibnamefont {Potter}},
  \bibinfo {author} {\bibfnamefont {I.-D.}\ \bibnamefont {Potirniche}},\ and\
  \bibinfo {author} {\bibfnamefont {A.}~\bibnamefont {Vishwanath}},\ }\href
  {https://doi.org/10.1103/PhysRevLett.118.030401} {\bibfield  {journal}
  {\bibinfo  {journal} {Phys. Rev. Lett.}\ }\textbf {\bibinfo {volume} {118}},\
  \bibinfo {pages} {030401} (\bibinfo {year} {2017})}\BibitemShut {NoStop}%
\bibitem [{\citenamefont {Ippoliti}\ \emph {et~al.}(2021)\citenamefont
  {Ippoliti}, \citenamefont {Kechedzhi}, \citenamefont {Moessner},
  \citenamefont {Sondhi},\ and\ \citenamefont {Khemani}}]{Ippoliti2021}%
  \BibitemOpen
  \bibfield  {author} {\bibinfo {author} {\bibfnamefont {M.}~\bibnamefont
  {Ippoliti}}, \bibinfo {author} {\bibfnamefont {K.}~\bibnamefont {Kechedzhi}},
  \bibinfo {author} {\bibfnamefont {R.}~\bibnamefont {Moessner}}, \bibinfo
  {author} {\bibfnamefont {S.~L.}\ \bibnamefont {Sondhi}},\ and\ \bibinfo
  {author} {\bibfnamefont {V.}~\bibnamefont {Khemani}},\ }\href
  {https://doi.org/10.1103/PRXQuantum.2.030346} {\bibfield  {journal} {\bibinfo
   {journal} {PRX Quantum}\ }\textbf {\bibinfo {volume} {2}},\ \bibinfo {pages}
  {030346} (\bibinfo {year} {2021})}\BibitemShut {NoStop}%
\bibitem [{\citenamefont {Nandkishore}\ and\ \citenamefont
  {Huse}(2015)}]{Huse2015}%
  \BibitemOpen
  \bibfield  {author} {\bibinfo {author} {\bibfnamefont {R.}~\bibnamefont
  {Nandkishore}}\ and\ \bibinfo {author} {\bibfnamefont {D.~A.}\ \bibnamefont
  {Huse}},\ }\href {https://doi.org/10.1146/annurev-conmatphys-031214-014726}
  {\bibfield  {journal} {\bibinfo  {journal} {Annual Review of Condensed Matter
  Physics}\ }\textbf {\bibinfo {volume} {6}},\ \bibinfo {pages} {15} (\bibinfo
  {year} {2015})},\ \Eprint
  {https://arxiv.org/abs/https://doi.org/10.1146/annurev-conmatphys-031214-014726}
  {https://doi.org/10.1146/annurev-conmatphys-031214-014726} \BibitemShut
  {NoStop}%
\bibitem [{\citenamefont {Imbrie}(2016)}]{Imbrie2016}%
  \BibitemOpen
  \bibfield  {author} {\bibinfo {author} {\bibfnamefont {J.~Z.}\ \bibnamefont
  {Imbrie}},\ }\href {https://doi.org/10.1007/s10955-016-1508-x} {\bibfield
  {journal} {\bibinfo  {journal} {Journal of Statistical Physics}\ }\textbf
  {\bibinfo {volume} {163}},\ \bibinfo {pages} {998} (\bibinfo {year}
  {2016})}\BibitemShut {NoStop}%
\bibitem [{\citenamefont {Ponte}\ \emph {et~al.}(2015)\citenamefont {Ponte},
  \citenamefont {Papi\ifmmode~\acute{c}\else \'{c}\fi{}}, \citenamefont
  {Huveneers},\ and\ \citenamefont {Abanin}}]{Ponte2015}%
  \BibitemOpen
  \bibfield  {author} {\bibinfo {author} {\bibfnamefont {P.}~\bibnamefont
  {Ponte}}, \bibinfo {author} {\bibfnamefont {Z.}~\bibnamefont
  {Papi\ifmmode~\acute{c}\else \'{c}\fi{}}}, \bibinfo {author} {\bibfnamefont
  {F.}~\bibnamefont {Huveneers}},\ and\ \bibinfo {author} {\bibfnamefont
  {D.~A.}\ \bibnamefont {Abanin}},\ }\href
  {https://doi.org/10.1103/PhysRevLett.114.140401} {\bibfield  {journal}
  {\bibinfo  {journal} {Phys. Rev. Lett.}\ }\textbf {\bibinfo {volume} {114}},\
  \bibinfo {pages} {140401} (\bibinfo {year} {2015})}\BibitemShut {NoStop}%
\bibitem [{\citenamefont {Faraday}(1831)}]{Faraday1831}%
  \BibitemOpen
  \bibfield  {author} {\bibinfo {author} {\bibfnamefont {M.}~\bibnamefont
  {Faraday}},\ }\href {http://www.jstor.org/stable/107936} {\bibfield
  {journal} {\bibinfo  {journal} {Philosophical Transactions of the Royal
  Society of London}\ }\textbf {\bibinfo {volume} {121}},\ \bibinfo {pages}
  {299} (\bibinfo {year} {1831})}\BibitemShut {NoStop}%
\bibitem [{\citenamefont {Lazarides}\ \emph {et~al.}(2015)\citenamefont
  {Lazarides}, \citenamefont {Das},\ and\ \citenamefont
  {Moessner}}]{Lazarides2015}%
  \BibitemOpen
  \bibfield  {author} {\bibinfo {author} {\bibfnamefont {A.}~\bibnamefont
  {Lazarides}}, \bibinfo {author} {\bibfnamefont {A.}~\bibnamefont {Das}},\
  and\ \bibinfo {author} {\bibfnamefont {R.}~\bibnamefont {Moessner}},\ }\href
  {https://doi.org/10.1103/PhysRevLett.115.030402} {\bibfield  {journal}
  {\bibinfo  {journal} {Phys. Rev. Lett.}\ }\textbf {\bibinfo {volume} {115}},\
  \bibinfo {pages} {030402} (\bibinfo {year} {2015})}\BibitemShut {NoStop}%
\bibitem [{\citenamefont {Abanin}\ \emph {et~al.}(2021)\citenamefont {Abanin},
  \citenamefont {Bardarson}, \citenamefont {{De Tomasi}}, \citenamefont
  {Gopalakrishnan}, \citenamefont {Khemani}, \citenamefont {Parameswaran},
  \citenamefont {Pollmann}, \citenamefont {Potter}, \citenamefont {Serbyn},\
  and\ \citenamefont {Vasseur}}]{Abanin2021}%
  \BibitemOpen
  \bibfield  {author} {\bibinfo {author} {\bibfnamefont {D.}~\bibnamefont
  {Abanin}}, \bibinfo {author} {\bibfnamefont {J.}~\bibnamefont {Bardarson}},
  \bibinfo {author} {\bibfnamefont {G.}~\bibnamefont {{De Tomasi}}}, \bibinfo
  {author} {\bibfnamefont {S.}~\bibnamefont {Gopalakrishnan}}, \bibinfo
  {author} {\bibfnamefont {V.}~\bibnamefont {Khemani}}, \bibinfo {author}
  {\bibfnamefont {S.}~\bibnamefont {Parameswaran}}, \bibinfo {author}
  {\bibfnamefont {F.}~\bibnamefont {Pollmann}}, \bibinfo {author}
  {\bibfnamefont {A.}~\bibnamefont {Potter}}, \bibinfo {author} {\bibfnamefont
  {M.}~\bibnamefont {Serbyn}},\ and\ \bibinfo {author} {\bibfnamefont
  {R.}~\bibnamefont {Vasseur}},\ }\href
  {https://doi.org/https://doi.org/10.1016/j.aop.2021.168415} {\bibfield
  {journal} {\bibinfo  {journal} {Annals of Physics}\ }\textbf {\bibinfo
  {volume} {427}},\ \bibinfo {pages} {168415} (\bibinfo {year}
  {2021})}\BibitemShut {NoStop}%
\bibitem [{\citenamefont {Morningstar}\ \emph {et~al.}(2022)\citenamefont
  {Morningstar}, \citenamefont {Colmenarez}, \citenamefont {Khemani},
  \citenamefont {Luitz},\ and\ \citenamefont {Huse}}]{Morningstar2021}%
  \BibitemOpen
  \bibfield  {author} {\bibinfo {author} {\bibfnamefont {A.}~\bibnamefont
  {Morningstar}}, \bibinfo {author} {\bibfnamefont {L.}~\bibnamefont
  {Colmenarez}}, \bibinfo {author} {\bibfnamefont {V.}~\bibnamefont {Khemani}},
  \bibinfo {author} {\bibfnamefont {D.~J.}\ \bibnamefont {Luitz}},\ and\
  \bibinfo {author} {\bibfnamefont {D.~A.}\ \bibnamefont {Huse}},\ }\href
  {https://doi.org/10.1103/PhysRevB.105.174205} {\bibfield  {journal} {\bibinfo
   {journal} {Phys. Rev. B}\ }\textbf {\bibinfo {volume} {105}},\ \bibinfo
  {pages} {174205} (\bibinfo {year} {2022})}\BibitemShut {NoStop}%
\bibitem [{\citenamefont {Sels}(2021)}]{Sels2021}%
  \BibitemOpen
  \bibfield  {author} {\bibinfo {author} {\bibfnamefont {D.}~\bibnamefont
  {Sels}},\ }\Eprint {https://arxiv.org/abs/arXiv:2108.10796}
  {arXiv:arXiv:2108.10796}  (\bibinfo {year} {2021})\BibitemShut {NoStop}%
\bibitem [{\citenamefont {Tu}\ \emph {et~al.}(2022)\citenamefont {Tu},
  \citenamefont {Vu},\ and\ \citenamefont {Das~Sarma}}]{Tu2022}%
  \BibitemOpen
  \bibfield  {author} {\bibinfo {author} {\bibfnamefont {Y.-T.}\ \bibnamefont
  {Tu}}, \bibinfo {author} {\bibfnamefont {D.}~\bibnamefont {Vu}},\ and\
  \bibinfo {author} {\bibfnamefont {S.}~\bibnamefont {Das~Sarma}},\ }\Eprint
  {https://arxiv.org/abs/arXiv:2207.05051} {arXiv:arXiv:2207.05051}  (\bibinfo
  {year} {2022})\BibitemShut {NoStop}%
\bibitem [{\citenamefont {Moessner}\ and\ \citenamefont
  {Sondhi}(2017)}]{Moessner2017}%
  \BibitemOpen
  \bibfield  {author} {\bibinfo {author} {\bibfnamefont {R.}~\bibnamefont
  {Moessner}}\ and\ \bibinfo {author} {\bibfnamefont {S.~L.}\ \bibnamefont
  {Sondhi}},\ }\href {https://doi.org/10.1038/nphys4106} {\bibfield  {journal}
  {\bibinfo  {journal} {Nature Physics}\ }\textbf {\bibinfo {volume} {13}},\
  \bibinfo {pages} {424} (\bibinfo {year} {2017})}\BibitemShut {NoStop}%
\bibitem [{\citenamefont {Harper}\ \emph {et~al.}(2020)\citenamefont {Harper},
  \citenamefont {Roy}, \citenamefont {Rudner},\ and\ \citenamefont
  {Sondhi}}]{Harper2020}%
  \BibitemOpen
  \bibfield  {author} {\bibinfo {author} {\bibfnamefont {F.}~\bibnamefont
  {Harper}}, \bibinfo {author} {\bibfnamefont {R.}~\bibnamefont {Roy}},
  \bibinfo {author} {\bibfnamefont {M.~S.}\ \bibnamefont {Rudner}},\ and\
  \bibinfo {author} {\bibfnamefont {S.}~\bibnamefont {Sondhi}},\ }\href
  {https://doi.org/10.1146/annurev-conmatphys-031218-013721} {\bibfield
  {journal} {\bibinfo  {journal} {Annual Review of Condensed Matter Physics}\
  }\textbf {\bibinfo {volume} {11}},\ \bibinfo {pages} {345} (\bibinfo {year}
  {2020})},\ \Eprint
  {https://arxiv.org/abs/https://doi.org/10.1146/annurev-conmatphys-031218-013721}
  {https://doi.org/10.1146/annurev-conmatphys-031218-013721} \BibitemShut
  {NoStop}%
\bibitem [{\citenamefont {Zhang}\ \emph {et~al.}(2022)\citenamefont {Zhang},
  \citenamefont {Jiang}, \citenamefont {Deng}, \citenamefont {Wang},
  \citenamefont {Chen}, \citenamefont {Zhang}, \citenamefont {Ren},
  \citenamefont {Dong}, \citenamefont {Xu}, \citenamefont {Gao}, \citenamefont
  {Jin}, \citenamefont {Zhu}, \citenamefont {Guo}, \citenamefont {Li},
  \citenamefont {Song}, \citenamefont {Gorshkov}, \citenamefont {Iadecola},
  \citenamefont {Liu}, \citenamefont {Gong}, \citenamefont {Wang},
  \citenamefont {Deng},\ and\ \citenamefont {Wang}}]{Zhang2022}%
  \BibitemOpen
  \bibfield  {author} {\bibinfo {author} {\bibfnamefont {X.}~\bibnamefont
  {Zhang}}, \bibinfo {author} {\bibfnamefont {W.}~\bibnamefont {Jiang}},
  \bibinfo {author} {\bibfnamefont {J.}~\bibnamefont {Deng}}, \bibinfo {author}
  {\bibfnamefont {K.}~\bibnamefont {Wang}}, \bibinfo {author} {\bibfnamefont
  {J.}~\bibnamefont {Chen}}, \bibinfo {author} {\bibfnamefont {P.}~\bibnamefont
  {Zhang}}, \bibinfo {author} {\bibfnamefont {W.}~\bibnamefont {Ren}}, \bibinfo
  {author} {\bibfnamefont {H.}~\bibnamefont {Dong}}, \bibinfo {author}
  {\bibfnamefont {S.}~\bibnamefont {Xu}}, \bibinfo {author} {\bibfnamefont
  {Y.}~\bibnamefont {Gao}}, \bibinfo {author} {\bibfnamefont {F.}~\bibnamefont
  {Jin}}, \bibinfo {author} {\bibfnamefont {X.}~\bibnamefont {Zhu}}, \bibinfo
  {author} {\bibfnamefont {Q.}~\bibnamefont {Guo}}, \bibinfo {author}
  {\bibfnamefont {H.}~\bibnamefont {Li}}, \bibinfo {author} {\bibfnamefont
  {C.}~\bibnamefont {Song}}, \bibinfo {author} {\bibfnamefont {A.~V.}\
  \bibnamefont {Gorshkov}}, \bibinfo {author} {\bibfnamefont {T.}~\bibnamefont
  {Iadecola}}, \bibinfo {author} {\bibfnamefont {F.}~\bibnamefont {Liu}},
  \bibinfo {author} {\bibfnamefont {Z.-X.}\ \bibnamefont {Gong}}, \bibinfo
  {author} {\bibfnamefont {Z.}~\bibnamefont {Wang}}, \bibinfo {author}
  {\bibfnamefont {D.-L.}\ \bibnamefont {Deng}},\ and\ \bibinfo {author}
  {\bibfnamefont {H.}~\bibnamefont {Wang}},\ }\href
  {https://doi.org/10.1038/s41586-022-04854-3} {\bibfield  {journal} {\bibinfo
  {journal} {Nature}\ }\textbf {\bibinfo {volume} {607}},\ \bibinfo {pages}
  {468} (\bibinfo {year} {2022})}\BibitemShut {NoStop}%
\bibitem [{\citenamefont {Loss}\ and\ \citenamefont
  {DiVincenzo}(1998)}]{Loss1998}%
  \BibitemOpen
  \bibfield  {author} {\bibinfo {author} {\bibfnamefont {D.}~\bibnamefont
  {Loss}}\ and\ \bibinfo {author} {\bibfnamefont {D.~P.}\ \bibnamefont
  {DiVincenzo}},\ }\href {https://doi.org/10.1103/PhysRevA.57.120} {\bibfield
  {journal} {\bibinfo  {journal} {Phys. Rev. A}\ }\textbf {\bibinfo {volume}
  {57}},\ \bibinfo {pages} {120} (\bibinfo {year} {1998})}\BibitemShut
  {NoStop}%
\bibitem [{\citenamefont {Mills}\ \emph {et~al.}(2022)\citenamefont {Mills},
  \citenamefont {Guinn}, \citenamefont {Gullans}, \citenamefont {Sigillito},
  \citenamefont {Feldman}, \citenamefont {Nielsen},\ and\ \citenamefont
  {Petta}}]{Mills2021}%
  \BibitemOpen
  \bibfield  {author} {\bibinfo {author} {\bibfnamefont {A.~R.}\ \bibnamefont
  {Mills}}, \bibinfo {author} {\bibfnamefont {C.~R.}\ \bibnamefont {Guinn}},
  \bibinfo {author} {\bibfnamefont {M.~J.}\ \bibnamefont {Gullans}}, \bibinfo
  {author} {\bibfnamefont {A.~J.}\ \bibnamefont {Sigillito}}, \bibinfo {author}
  {\bibfnamefont {M.~M.}\ \bibnamefont {Feldman}}, \bibinfo {author}
  {\bibfnamefont {E.}~\bibnamefont {Nielsen}},\ and\ \bibinfo {author}
  {\bibfnamefont {J.~R.}\ \bibnamefont {Petta}},\ }\href
  {https://doi.org/10.1126/sciadv.abn5130} {\bibfield  {journal} {\bibinfo
  {journal} {Science Advances}\ }\textbf {\bibinfo {volume} {8}},\ \bibinfo
  {pages} {eabn5130} (\bibinfo {year} {2022})},\ \Eprint
  {https://arxiv.org/abs/https://www.science.org/doi/pdf/10.1126/sciadv.abn5130}
  {https://www.science.org/doi/pdf/10.1126/sciadv.abn5130} \BibitemShut
  {NoStop}%
\bibitem [{\citenamefont {Noiri}\ \emph {et~al.}(2022)\citenamefont {Noiri},
  \citenamefont {Takeda}, \citenamefont {Nakajima}, \citenamefont {Kobayashi},
  \citenamefont {Sammak}, \citenamefont {Scappucci},\ and\ \citenamefont
  {Tarucha}}]{Noiri2022}%
  \BibitemOpen
  \bibfield  {author} {\bibinfo {author} {\bibfnamefont {A.}~\bibnamefont
  {Noiri}}, \bibinfo {author} {\bibfnamefont {K.}~\bibnamefont {Takeda}},
  \bibinfo {author} {\bibfnamefont {T.}~\bibnamefont {Nakajima}}, \bibinfo
  {author} {\bibfnamefont {T.}~\bibnamefont {Kobayashi}}, \bibinfo {author}
  {\bibfnamefont {A.}~\bibnamefont {Sammak}}, \bibinfo {author} {\bibfnamefont
  {G.}~\bibnamefont {Scappucci}},\ and\ \bibinfo {author} {\bibfnamefont
  {S.}~\bibnamefont {Tarucha}},\ }\href
  {https://doi.org/10.1038/s41586-021-04182-y} {\bibfield  {journal} {\bibinfo
  {journal} {Nature}\ }\textbf {\bibinfo {volume} {601}},\ \bibinfo {pages}
  {338} (\bibinfo {year} {2022})}\BibitemShut {NoStop}%
\bibitem [{\citenamefont {Philips}\ \emph {et~al.}(2022)\citenamefont
  {Philips}, \citenamefont {M\k{a}dzik}, \citenamefont {Amitonov},
  \citenamefont {de~Snoo}, \citenamefont {Russ}, \citenamefont {Kalhor},
  \citenamefont {Volk}, \citenamefont {Lawrie}, \citenamefont {Brousse},
  \citenamefont {Tryputen}, \citenamefont {Wuetz}, \citenamefont {Sammak},
  \citenamefont {Veldhorst}, \citenamefont {Scappucci},\ and\ \citenamefont
  {Vandersypen}}]{Philips2022}%
  \BibitemOpen
  \bibfield  {author} {\bibinfo {author} {\bibfnamefont {S.~G.~J.}\
  \bibnamefont {Philips}}, \bibinfo {author} {\bibfnamefont {M.~T.}\
  \bibnamefont {M\k{a}dzik}}, \bibinfo {author} {\bibfnamefont {S.~V.}\
  \bibnamefont {Amitonov}}, \bibinfo {author} {\bibfnamefont {S.~L.}\
  \bibnamefont {de~Snoo}}, \bibinfo {author} {\bibfnamefont {M.}~\bibnamefont
  {Russ}}, \bibinfo {author} {\bibfnamefont {N.}~\bibnamefont {Kalhor}},
  \bibinfo {author} {\bibfnamefont {C.}~\bibnamefont {Volk}}, \bibinfo {author}
  {\bibfnamefont {W.~I.~L.}\ \bibnamefont {Lawrie}}, \bibinfo {author}
  {\bibfnamefont {D.}~\bibnamefont {Brousse}}, \bibinfo {author} {\bibfnamefont
  {L.}~\bibnamefont {Tryputen}}, \bibinfo {author} {\bibfnamefont {B.~P.}\
  \bibnamefont {Wuetz}}, \bibinfo {author} {\bibfnamefont {A.}~\bibnamefont
  {Sammak}}, \bibinfo {author} {\bibfnamefont {M.}~\bibnamefont {Veldhorst}},
  \bibinfo {author} {\bibfnamefont {G.}~\bibnamefont {Scappucci}},\ and\
  \bibinfo {author} {\bibfnamefont {L.~M.~K.}\ \bibnamefont {Vandersypen}},\
  }\href {https://doi.org/10.1038/s41586-022-05117-x} {\bibfield  {journal}
  {\bibinfo  {journal} {Nature}\ }\textbf {\bibinfo {volume} {609}},\ \bibinfo
  {pages} {919} (\bibinfo {year} {2022})}\BibitemShut {NoStop}%
\bibitem [{\citenamefont {Fowler}\ \emph {et~al.}(2012)\citenamefont {Fowler},
  \citenamefont {Mariantoni}, \citenamefont {Martinis},\ and\ \citenamefont
  {Cleland}}]{Fowler2012}%
  \BibitemOpen
  \bibfield  {author} {\bibinfo {author} {\bibfnamefont {A.~G.}\ \bibnamefont
  {Fowler}}, \bibinfo {author} {\bibfnamefont {M.}~\bibnamefont {Mariantoni}},
  \bibinfo {author} {\bibfnamefont {J.~M.}\ \bibnamefont {Martinis}},\ and\
  \bibinfo {author} {\bibfnamefont {A.~N.}\ \bibnamefont {Cleland}},\
  }\href@noop {} {\bibfield  {journal} {\bibinfo  {journal} {Physical Review
  A}\ }\textbf {\bibinfo {volume} {86}},\ \bibinfo {pages} {032324} (\bibinfo
  {year} {2012})}\BibitemShut {NoStop}%
\bibitem [{\citenamefont {Borsoi}\ \emph {et~al.}(2022)\citenamefont {Borsoi},
  \citenamefont {Hendrickx}, \citenamefont {John}, \citenamefont {Motz},
  \citenamefont {van Riggelen}, \citenamefont {Sammak}, \citenamefont
  {de~Snoo}, \citenamefont {Scappucci},\ and\ \citenamefont
  {Veldhorst}}]{Borsoi2022}%
  \BibitemOpen
  \bibfield  {author} {\bibinfo {author} {\bibfnamefont {F.}~\bibnamefont
  {Borsoi}}, \bibinfo {author} {\bibfnamefont {N.~W.}\ \bibnamefont
  {Hendrickx}}, \bibinfo {author} {\bibfnamefont {V.}~\bibnamefont {John}},
  \bibinfo {author} {\bibfnamefont {S.}~\bibnamefont {Motz}}, \bibinfo {author}
  {\bibfnamefont {F.}~\bibnamefont {van Riggelen}}, \bibinfo {author}
  {\bibfnamefont {A.}~\bibnamefont {Sammak}}, \bibinfo {author} {\bibfnamefont
  {S.~L.}\ \bibnamefont {de~Snoo}}, \bibinfo {author} {\bibfnamefont
  {G.}~\bibnamefont {Scappucci}},\ and\ \bibinfo {author} {\bibfnamefont
  {M.}~\bibnamefont {Veldhorst}},\ }\Eprint
  {https://arxiv.org/abs/arXiv:2209.06609} {arXiv:arXiv:2209.06609}  (\bibinfo
  {year} {2022})\BibitemShut {NoStop}%
\bibitem [{\citenamefont {Mi}\ \emph {et~al.}(2022)\citenamefont {Mi},
  \citenamefont {Ippoliti}, \citenamefont {Quintana}, \citenamefont {Greene},
  \citenamefont {Chen}, \citenamefont {Gross}, \citenamefont {Arute},
  \citenamefont {Arya}, \citenamefont {Atalaya}, \citenamefont {Babbush},
  \citenamefont {Bardin}, \citenamefont {Basso}, \citenamefont {Bengtsson},
  \citenamefont {Bilmes}, \citenamefont {Bourassa}, \citenamefont {Brill},
  \citenamefont {Broughton}, \citenamefont {Buckley}, \citenamefont {Buell},
  \citenamefont {Burkett}, \citenamefont {Bushnell}, \citenamefont {Chiaro},
  \citenamefont {Collins}, \citenamefont {Courtney}, \citenamefont {Debroy},
  \citenamefont {Demura}, \citenamefont {Derk}, \citenamefont {Dunsworth},
  \citenamefont {Eppens}, \citenamefont {Erickson}, \citenamefont {Farhi},
  \citenamefont {Fowler}, \citenamefont {Foxen}, \citenamefont {Gidney},
  \citenamefont {Giustina}, \citenamefont {Harrigan}, \citenamefont
  {Harrington}, \citenamefont {Hilton}, \citenamefont {Ho}, \citenamefont
  {Hong}, \citenamefont {Huang}, \citenamefont {Huff}, \citenamefont {Huggins},
  \citenamefont {Ioffe}, \citenamefont {Isakov}, \citenamefont {Iveland},
  \citenamefont {Jeffrey}, \citenamefont {Jiang}, \citenamefont {Jones},
  \citenamefont {Kafri}, \citenamefont {Khattar}, \citenamefont {Kim},
  \citenamefont {Kitaev}, \citenamefont {Klimov}, \citenamefont {Korotkov},
  \citenamefont {Kostritsa}, \citenamefont {Landhuis}, \citenamefont {Laptev},
  \citenamefont {Lee}, \citenamefont {Lee}, \citenamefont {Locharla},
  \citenamefont {Lucero}, \citenamefont {Martin}, \citenamefont {McClean},
  \citenamefont {McCourt}, \citenamefont {McEwen}, \citenamefont {Miao},
  \citenamefont {Mohseni}, \citenamefont {Montazeri}, \citenamefont
  {Mruczkiewicz}, \citenamefont {Naaman}, \citenamefont {Neeley}, \citenamefont
  {Neill}, \citenamefont {Newman}, \citenamefont {Niu}, \citenamefont
  {O'Brien}, \citenamefont {Opremcak}, \citenamefont {Ostby}, \citenamefont
  {Pato}, \citenamefont {Petukhov}, \citenamefont {Rubin}, \citenamefont
  {Sank}, \citenamefont {Satzinger}, \citenamefont {Shvarts}, \citenamefont
  {Su}, \citenamefont {Strain}, \citenamefont {Szalay}, \citenamefont
  {Trevithick}, \citenamefont {Villalonga}, \citenamefont {White},
  \citenamefont {Yao}, \citenamefont {Yeh}, \citenamefont {Yoo}, \citenamefont
  {Zalcman}, \citenamefont {Neven}, \citenamefont {Boixo}, \citenamefont
  {Smelyanskiy}, \citenamefont {Megrant}, \citenamefont {Kelly}, \citenamefont
  {Chen}, \citenamefont {Sondhi}, \citenamefont {Moessner}, \citenamefont
  {Kechedzhi}, \citenamefont {Khemani},\ and\ \citenamefont
  {Roushan}}]{Mi2022}%
  \BibitemOpen
  \bibfield  {author} {\bibinfo {author} {\bibfnamefont {X.}~\bibnamefont
  {Mi}}, \bibinfo {author} {\bibfnamefont {M.}~\bibnamefont {Ippoliti}},
  \bibinfo {author} {\bibfnamefont {C.}~\bibnamefont {Quintana}}, \bibinfo
  {author} {\bibfnamefont {A.}~\bibnamefont {Greene}}, \bibinfo {author}
  {\bibfnamefont {Z.}~\bibnamefont {Chen}}, \bibinfo {author} {\bibfnamefont
  {J.}~\bibnamefont {Gross}}, \bibinfo {author} {\bibfnamefont
  {F.}~\bibnamefont {Arute}}, \bibinfo {author} {\bibfnamefont
  {K.}~\bibnamefont {Arya}}, \bibinfo {author} {\bibfnamefont {J.}~\bibnamefont
  {Atalaya}}, \bibinfo {author} {\bibfnamefont {R.}~\bibnamefont {Babbush}},
  \bibinfo {author} {\bibfnamefont {J.~C.}\ \bibnamefont {Bardin}}, \bibinfo
  {author} {\bibfnamefont {J.}~\bibnamefont {Basso}}, \bibinfo {author}
  {\bibfnamefont {A.}~\bibnamefont {Bengtsson}}, \bibinfo {author}
  {\bibfnamefont {A.}~\bibnamefont {Bilmes}}, \bibinfo {author} {\bibfnamefont
  {A.}~\bibnamefont {Bourassa}}, \bibinfo {author} {\bibfnamefont
  {L.}~\bibnamefont {Brill}}, \bibinfo {author} {\bibfnamefont
  {M.}~\bibnamefont {Broughton}}, \bibinfo {author} {\bibfnamefont {B.~B.}\
  \bibnamefont {Buckley}}, \bibinfo {author} {\bibfnamefont {D.~A.}\
  \bibnamefont {Buell}}, \bibinfo {author} {\bibfnamefont {B.}~\bibnamefont
  {Burkett}}, \bibinfo {author} {\bibfnamefont {N.}~\bibnamefont {Bushnell}},
  \bibinfo {author} {\bibfnamefont {B.}~\bibnamefont {Chiaro}}, \bibinfo
  {author} {\bibfnamefont {R.}~\bibnamefont {Collins}}, \bibinfo {author}
  {\bibfnamefont {W.}~\bibnamefont {Courtney}}, \bibinfo {author}
  {\bibfnamefont {D.}~\bibnamefont {Debroy}}, \bibinfo {author} {\bibfnamefont
  {S.}~\bibnamefont {Demura}}, \bibinfo {author} {\bibfnamefont {A.~R.}\
  \bibnamefont {Derk}}, \bibinfo {author} {\bibfnamefont {A.}~\bibnamefont
  {Dunsworth}}, \bibinfo {author} {\bibfnamefont {D.}~\bibnamefont {Eppens}},
  \bibinfo {author} {\bibfnamefont {C.}~\bibnamefont {Erickson}}, \bibinfo
  {author} {\bibfnamefont {E.}~\bibnamefont {Farhi}}, \bibinfo {author}
  {\bibfnamefont {A.~G.}\ \bibnamefont {Fowler}}, \bibinfo {author}
  {\bibfnamefont {B.}~\bibnamefont {Foxen}}, \bibinfo {author} {\bibfnamefont
  {C.}~\bibnamefont {Gidney}}, \bibinfo {author} {\bibfnamefont
  {M.}~\bibnamefont {Giustina}}, \bibinfo {author} {\bibfnamefont {M.~P.}\
  \bibnamefont {Harrigan}}, \bibinfo {author} {\bibfnamefont {S.~D.}\
  \bibnamefont {Harrington}}, \bibinfo {author} {\bibfnamefont
  {J.}~\bibnamefont {Hilton}}, \bibinfo {author} {\bibfnamefont
  {A.}~\bibnamefont {Ho}}, \bibinfo {author} {\bibfnamefont {S.}~\bibnamefont
  {Hong}}, \bibinfo {author} {\bibfnamefont {T.}~\bibnamefont {Huang}},
  \bibinfo {author} {\bibfnamefont {A.}~\bibnamefont {Huff}}, \bibinfo {author}
  {\bibfnamefont {W.~J.}\ \bibnamefont {Huggins}}, \bibinfo {author}
  {\bibfnamefont {L.~B.}\ \bibnamefont {Ioffe}}, \bibinfo {author}
  {\bibfnamefont {S.~V.}\ \bibnamefont {Isakov}}, \bibinfo {author}
  {\bibfnamefont {J.}~\bibnamefont {Iveland}}, \bibinfo {author} {\bibfnamefont
  {E.}~\bibnamefont {Jeffrey}}, \bibinfo {author} {\bibfnamefont
  {Z.}~\bibnamefont {Jiang}}, \bibinfo {author} {\bibfnamefont
  {C.}~\bibnamefont {Jones}}, \bibinfo {author} {\bibfnamefont
  {D.}~\bibnamefont {Kafri}}, \bibinfo {author} {\bibfnamefont
  {T.}~\bibnamefont {Khattar}}, \bibinfo {author} {\bibfnamefont
  {S.}~\bibnamefont {Kim}}, \bibinfo {author} {\bibfnamefont {A.}~\bibnamefont
  {Kitaev}}, \bibinfo {author} {\bibfnamefont {P.~V.}\ \bibnamefont {Klimov}},
  \bibinfo {author} {\bibfnamefont {A.~N.}\ \bibnamefont {Korotkov}}, \bibinfo
  {author} {\bibfnamefont {F.}~\bibnamefont {Kostritsa}}, \bibinfo {author}
  {\bibfnamefont {D.}~\bibnamefont {Landhuis}}, \bibinfo {author}
  {\bibfnamefont {P.}~\bibnamefont {Laptev}}, \bibinfo {author} {\bibfnamefont
  {J.}~\bibnamefont {Lee}}, \bibinfo {author} {\bibfnamefont {K.}~\bibnamefont
  {Lee}}, \bibinfo {author} {\bibfnamefont {A.}~\bibnamefont {Locharla}},
  \bibinfo {author} {\bibfnamefont {E.}~\bibnamefont {Lucero}}, \bibinfo
  {author} {\bibfnamefont {O.}~\bibnamefont {Martin}}, \bibinfo {author}
  {\bibfnamefont {J.~R.}\ \bibnamefont {McClean}}, \bibinfo {author}
  {\bibfnamefont {T.}~\bibnamefont {McCourt}}, \bibinfo {author} {\bibfnamefont
  {M.}~\bibnamefont {McEwen}}, \bibinfo {author} {\bibfnamefont {K.~C.}\
  \bibnamefont {Miao}}, \bibinfo {author} {\bibfnamefont {M.}~\bibnamefont
  {Mohseni}}, \bibinfo {author} {\bibfnamefont {S.}~\bibnamefont {Montazeri}},
  \bibinfo {author} {\bibfnamefont {W.}~\bibnamefont {Mruczkiewicz}}, \bibinfo
  {author} {\bibfnamefont {O.}~\bibnamefont {Naaman}}, \bibinfo {author}
  {\bibfnamefont {M.}~\bibnamefont {Neeley}}, \bibinfo {author} {\bibfnamefont
  {C.}~\bibnamefont {Neill}}, \bibinfo {author} {\bibfnamefont
  {M.}~\bibnamefont {Newman}}, \bibinfo {author} {\bibfnamefont {M.~Y.}\
  \bibnamefont {Niu}}, \bibinfo {author} {\bibfnamefont {T.~E.}\ \bibnamefont
  {O'Brien}}, \bibinfo {author} {\bibfnamefont {A.}~\bibnamefont {Opremcak}},
  \bibinfo {author} {\bibfnamefont {E.}~\bibnamefont {Ostby}}, \bibinfo
  {author} {\bibfnamefont {B.}~\bibnamefont {Pato}}, \bibinfo {author}
  {\bibfnamefont {A.}~\bibnamefont {Petukhov}}, \bibinfo {author}
  {\bibfnamefont {N.~C.}\ \bibnamefont {Rubin}}, \bibinfo {author}
  {\bibfnamefont {D.}~\bibnamefont {Sank}}, \bibinfo {author} {\bibfnamefont
  {K.~J.}\ \bibnamefont {Satzinger}}, \bibinfo {author} {\bibfnamefont
  {V.}~\bibnamefont {Shvarts}}, \bibinfo {author} {\bibfnamefont
  {Y.}~\bibnamefont {Su}}, \bibinfo {author} {\bibfnamefont {D.}~\bibnamefont
  {Strain}}, \bibinfo {author} {\bibfnamefont {M.}~\bibnamefont {Szalay}},
  \bibinfo {author} {\bibfnamefont {M.~D.}\ \bibnamefont {Trevithick}},
  \bibinfo {author} {\bibfnamefont {B.}~\bibnamefont {Villalonga}}, \bibinfo
  {author} {\bibfnamefont {T.}~\bibnamefont {White}}, \bibinfo {author}
  {\bibfnamefont {Z.~J.}\ \bibnamefont {Yao}}, \bibinfo {author} {\bibfnamefont
  {P.}~\bibnamefont {Yeh}}, \bibinfo {author} {\bibfnamefont {J.}~\bibnamefont
  {Yoo}}, \bibinfo {author} {\bibfnamefont {A.}~\bibnamefont {Zalcman}},
  \bibinfo {author} {\bibfnamefont {H.}~\bibnamefont {Neven}}, \bibinfo
  {author} {\bibfnamefont {S.}~\bibnamefont {Boixo}}, \bibinfo {author}
  {\bibfnamefont {V.}~\bibnamefont {Smelyanskiy}}, \bibinfo {author}
  {\bibfnamefont {A.}~\bibnamefont {Megrant}}, \bibinfo {author} {\bibfnamefont
  {J.}~\bibnamefont {Kelly}}, \bibinfo {author} {\bibfnamefont
  {Y.}~\bibnamefont {Chen}}, \bibinfo {author} {\bibfnamefont {S.~L.}\
  \bibnamefont {Sondhi}}, \bibinfo {author} {\bibfnamefont {R.}~\bibnamefont
  {Moessner}}, \bibinfo {author} {\bibfnamefont {K.}~\bibnamefont {Kechedzhi}},
  \bibinfo {author} {\bibfnamefont {V.}~\bibnamefont {Khemani}},\ and\ \bibinfo
  {author} {\bibfnamefont {P.}~\bibnamefont {Roushan}},\ }\href
  {https://doi.org/10.1038/s41586-021-04257-w} {\bibfield  {journal} {\bibinfo
  {journal} {Nature}\ }\textbf {\bibinfo {volume} {601}},\ \bibinfo {pages}
  {531} (\bibinfo {year} {2022})}\BibitemShut {NoStop}%
\bibitem [{\citenamefont {Zhang}\ \emph {et~al.}(2017)\citenamefont {Zhang},
  \citenamefont {Hess}, \citenamefont {Kyprianidis}, \citenamefont {Becker},
  \citenamefont {Lee}, \citenamefont {Smith}, \citenamefont {Pagano},
  \citenamefont {Potirniche}, \citenamefont {Potter}, \citenamefont
  {Vishwanath}, \citenamefont {Yao},\ and\ \citenamefont {Monroe}}]{Zhang2017}%
  \BibitemOpen
  \bibfield  {author} {\bibinfo {author} {\bibfnamefont {J.}~\bibnamefont
  {Zhang}}, \bibinfo {author} {\bibfnamefont {P.~W.}\ \bibnamefont {Hess}},
  \bibinfo {author} {\bibfnamefont {A.}~\bibnamefont {Kyprianidis}}, \bibinfo
  {author} {\bibfnamefont {P.}~\bibnamefont {Becker}}, \bibinfo {author}
  {\bibfnamefont {A.}~\bibnamefont {Lee}}, \bibinfo {author} {\bibfnamefont
  {J.}~\bibnamefont {Smith}}, \bibinfo {author} {\bibfnamefont
  {G.}~\bibnamefont {Pagano}}, \bibinfo {author} {\bibfnamefont {I.~D.}\
  \bibnamefont {Potirniche}}, \bibinfo {author} {\bibfnamefont {A.~C.}\
  \bibnamefont {Potter}}, \bibinfo {author} {\bibfnamefont {A.}~\bibnamefont
  {Vishwanath}}, \bibinfo {author} {\bibfnamefont {N.~Y.}\ \bibnamefont
  {Yao}},\ and\ \bibinfo {author} {\bibfnamefont {C.}~\bibnamefont {Monroe}},\
  }\href {https://doi.org/10.1038/nature21413} {\bibfield  {journal} {\bibinfo
  {journal} {Nature}\ }\textbf {\bibinfo {volume} {543}},\ \bibinfo {pages}
  {217} (\bibinfo {year} {2017})}\BibitemShut {NoStop}%
\bibitem [{\citenamefont {Rovny}\ \emph {et~al.}(2018)\citenamefont {Rovny},
  \citenamefont {Blum},\ and\ \citenamefont {Barrett}}]{Rovny2018}%
  \BibitemOpen
  \bibfield  {author} {\bibinfo {author} {\bibfnamefont {J.}~\bibnamefont
  {Rovny}}, \bibinfo {author} {\bibfnamefont {R.~L.}\ \bibnamefont {Blum}},\
  and\ \bibinfo {author} {\bibfnamefont {S.~E.}\ \bibnamefont {Barrett}},\
  }\href {https://doi.org/10.1103/PhysRevB.97.184301} {\bibfield  {journal}
  {\bibinfo  {journal} {Phys. Rev. B}\ }\textbf {\bibinfo {volume} {97}},\
  \bibinfo {pages} {184301} (\bibinfo {year} {2018})}\BibitemShut {NoStop}%
\bibitem [{\citenamefont {Choi}\ \emph {et~al.}(2017)\citenamefont {Choi},
  \citenamefont {Choi}, \citenamefont {Landig}, \citenamefont {Kucsko},
  \citenamefont {Zhou}, \citenamefont {Isoya}, \citenamefont {Jelezko},
  \citenamefont {Onoda}, \citenamefont {Sumiya}, \citenamefont {Khemani},
  \citenamefont {von Keyserlingk}, \citenamefont {Yao}, \citenamefont
  {Demler},\ and\ \citenamefont {Lukin}}]{Choi2017}%
  \BibitemOpen
  \bibfield  {author} {\bibinfo {author} {\bibfnamefont {S.}~\bibnamefont
  {Choi}}, \bibinfo {author} {\bibfnamefont {J.}~\bibnamefont {Choi}}, \bibinfo
  {author} {\bibfnamefont {R.}~\bibnamefont {Landig}}, \bibinfo {author}
  {\bibfnamefont {G.}~\bibnamefont {Kucsko}}, \bibinfo {author} {\bibfnamefont
  {H.}~\bibnamefont {Zhou}}, \bibinfo {author} {\bibfnamefont {J.}~\bibnamefont
  {Isoya}}, \bibinfo {author} {\bibfnamefont {F.}~\bibnamefont {Jelezko}},
  \bibinfo {author} {\bibfnamefont {S.}~\bibnamefont {Onoda}}, \bibinfo
  {author} {\bibfnamefont {H.}~\bibnamefont {Sumiya}}, \bibinfo {author}
  {\bibfnamefont {V.}~\bibnamefont {Khemani}}, \bibinfo {author} {\bibfnamefont
  {C.}~\bibnamefont {von Keyserlingk}}, \bibinfo {author} {\bibfnamefont
  {N.~Y.}\ \bibnamefont {Yao}}, \bibinfo {author} {\bibfnamefont
  {E.}~\bibnamefont {Demler}},\ and\ \bibinfo {author} {\bibfnamefont {M.~D.}\
  \bibnamefont {Lukin}},\ }\href {https://doi.org/10.1038/nature21426}
  {\bibfield  {journal} {\bibinfo  {journal} {Nature}\ }\textbf {\bibinfo
  {volume} {543}},\ \bibinfo {pages} {221} (\bibinfo {year}
  {2017})}\BibitemShut {NoStop}%
\bibitem [{\citenamefont {Barnes}\ \emph {et~al.}(2019)\citenamefont {Barnes},
  \citenamefont {Nichol},\ and\ \citenamefont {Economou}}]{Barnes2019}%
  \BibitemOpen
  \bibfield  {author} {\bibinfo {author} {\bibfnamefont {E.}~\bibnamefont
  {Barnes}}, \bibinfo {author} {\bibfnamefont {J.~M.}\ \bibnamefont {Nichol}},\
  and\ \bibinfo {author} {\bibfnamefont {S.~E.}\ \bibnamefont {Economou}},\
  }\href {https://doi.org/10.1103/PhysRevB.99.035311} {\bibfield  {journal}
  {\bibinfo  {journal} {Phys. Rev. B}\ }\textbf {\bibinfo {volume} {99}},\
  \bibinfo {pages} {035311} (\bibinfo {year} {2019})}\BibitemShut {NoStop}%
\bibitem [{\citenamefont {Sigillito}\ \emph {et~al.}(2019)\citenamefont
  {Sigillito}, \citenamefont {Loy}, \citenamefont {Zajac}, \citenamefont
  {Gullans}, \citenamefont {Edge},\ and\ \citenamefont
  {Petta}}]{Sigillito2019}%
  \BibitemOpen
  \bibfield  {author} {\bibinfo {author} {\bibfnamefont {A.~J.}\ \bibnamefont
  {Sigillito}}, \bibinfo {author} {\bibfnamefont {J.~C.}\ \bibnamefont {Loy}},
  \bibinfo {author} {\bibfnamefont {D.~M.}\ \bibnamefont {Zajac}}, \bibinfo
  {author} {\bibfnamefont {M.~J.}\ \bibnamefont {Gullans}}, \bibinfo {author}
  {\bibfnamefont {L.~F.}\ \bibnamefont {Edge}},\ and\ \bibinfo {author}
  {\bibfnamefont {J.~R.}\ \bibnamefont {Petta}},\ }\href
  {https://doi.org/10.1103/PhysRevApplied.11.061006} {\bibfield  {journal}
  {\bibinfo  {journal} {Phys. Rev. Appl.}\ }\textbf {\bibinfo {volume} {11}},\
  \bibinfo {pages} {061006(R)} (\bibinfo {year} {2019})}\BibitemShut {NoStop}%
\bibitem [{\citenamefont {Lawrie}\ \emph {et~al.}(2020)\citenamefont {Lawrie},
  \citenamefont {Eenink}, \citenamefont {Hendrickx}, \citenamefont {Boter},
  \citenamefont {Petit}, \citenamefont {Amitonov}, \citenamefont {Lodari},
  \citenamefont {Paquelet~Wuetz}, \citenamefont {Volk}, \citenamefont
  {Philips}, \citenamefont {Droulers}, \citenamefont {Kalhor}, \citenamefont
  {van Riggelen}, \citenamefont {Brousse}, \citenamefont {Sammak},
  \citenamefont {Vandersypen}, \citenamefont {Scappucci},\ and\ \citenamefont
  {Veldhorst}}]{Lawrie2020}%
  \BibitemOpen
  \bibfield  {author} {\bibinfo {author} {\bibfnamefont {W.~I.~L.}\
  \bibnamefont {Lawrie}}, \bibinfo {author} {\bibfnamefont {H.~G.~J.}\
  \bibnamefont {Eenink}}, \bibinfo {author} {\bibfnamefont {N.~W.}\
  \bibnamefont {Hendrickx}}, \bibinfo {author} {\bibfnamefont {J.~M.}\
  \bibnamefont {Boter}}, \bibinfo {author} {\bibfnamefont {L.}~\bibnamefont
  {Petit}}, \bibinfo {author} {\bibfnamefont {S.~V.}\ \bibnamefont {Amitonov}},
  \bibinfo {author} {\bibfnamefont {M.}~\bibnamefont {Lodari}}, \bibinfo
  {author} {\bibfnamefont {B.}~\bibnamefont {Paquelet~Wuetz}}, \bibinfo
  {author} {\bibfnamefont {C.}~\bibnamefont {Volk}}, \bibinfo {author}
  {\bibfnamefont {S.~G.~J.}\ \bibnamefont {Philips}}, \bibinfo {author}
  {\bibfnamefont {G.}~\bibnamefont {Droulers}}, \bibinfo {author}
  {\bibfnamefont {N.}~\bibnamefont {Kalhor}}, \bibinfo {author} {\bibfnamefont
  {F.}~\bibnamefont {van Riggelen}}, \bibinfo {author} {\bibfnamefont
  {D.}~\bibnamefont {Brousse}}, \bibinfo {author} {\bibfnamefont
  {A.}~\bibnamefont {Sammak}}, \bibinfo {author} {\bibfnamefont {L.~M.~K.}\
  \bibnamefont {Vandersypen}}, \bibinfo {author} {\bibfnamefont
  {G.}~\bibnamefont {Scappucci}},\ and\ \bibinfo {author} {\bibfnamefont
  {M.}~\bibnamefont {Veldhorst}},\ }\href {https://doi.org/10.1063/5.0002013}
  {\bibfield  {journal} {\bibinfo  {journal} {Applied Physics Letters}\
  }\textbf {\bibinfo {volume} {116}},\ \bibinfo {pages} {080501} (\bibinfo
  {year} {2020})},\ \Eprint
  {https://arxiv.org/abs/https://doi.org/10.1063/5.0002013}
  {https://doi.org/10.1063/5.0002013} \BibitemShut {NoStop}%
\bibitem [{\citenamefont {Reed}\ \emph {et~al.}(2016)\citenamefont {Reed},
  \citenamefont {Maune}, \citenamefont {Andrews}, \citenamefont {Borselli},
  \citenamefont {Eng}, \citenamefont {Jura}, \citenamefont {Kiselev},
  \citenamefont {Ladd}, \citenamefont {Merkel}, \citenamefont {Milosavljevic},
  \citenamefont {Pritchett}, \citenamefont {Rakher}, \citenamefont {Ross},
  \citenamefont {Schmitz}, \citenamefont {Smith}, \citenamefont {Wright},
  \citenamefont {Gyure},\ and\ \citenamefont {Hunter}}]{Reed2016}%
  \BibitemOpen
  \bibfield  {author} {\bibinfo {author} {\bibfnamefont {M.~D.}\ \bibnamefont
  {Reed}}, \bibinfo {author} {\bibfnamefont {B.~M.}\ \bibnamefont {Maune}},
  \bibinfo {author} {\bibfnamefont {R.~W.}\ \bibnamefont {Andrews}}, \bibinfo
  {author} {\bibfnamefont {M.~G.}\ \bibnamefont {Borselli}}, \bibinfo {author}
  {\bibfnamefont {K.}~\bibnamefont {Eng}}, \bibinfo {author} {\bibfnamefont
  {M.~P.}\ \bibnamefont {Jura}}, \bibinfo {author} {\bibfnamefont {A.~A.}\
  \bibnamefont {Kiselev}}, \bibinfo {author} {\bibfnamefont {T.~D.}\
  \bibnamefont {Ladd}}, \bibinfo {author} {\bibfnamefont {S.~T.}\ \bibnamefont
  {Merkel}}, \bibinfo {author} {\bibfnamefont {I.}~\bibnamefont
  {Milosavljevic}}, \bibinfo {author} {\bibfnamefont {E.~J.}\ \bibnamefont
  {Pritchett}}, \bibinfo {author} {\bibfnamefont {M.~T.}\ \bibnamefont
  {Rakher}}, \bibinfo {author} {\bibfnamefont {R.~S.}\ \bibnamefont {Ross}},
  \bibinfo {author} {\bibfnamefont {A.~E.}\ \bibnamefont {Schmitz}}, \bibinfo
  {author} {\bibfnamefont {A.}~\bibnamefont {Smith}}, \bibinfo {author}
  {\bibfnamefont {J.~A.}\ \bibnamefont {Wright}}, \bibinfo {author}
  {\bibfnamefont {M.~F.}\ \bibnamefont {Gyure}},\ and\ \bibinfo {author}
  {\bibfnamefont {A.~T.}\ \bibnamefont {Hunter}},\ }\href
  {https://doi.org/10.1103/PhysRevLett.116.110402} {\bibfield  {journal}
  {\bibinfo  {journal} {Phys. Rev. Lett.}\ }\textbf {\bibinfo {volume} {116}},\
  \bibinfo {pages} {110402} (\bibinfo {year} {2016})}\BibitemShut {NoStop}%
\bibitem [{\citenamefont {Connors}\ \emph {et~al.}(2022)\citenamefont
  {Connors}, \citenamefont {Nelson}, \citenamefont {Edge},\ and\ \citenamefont
  {Nichol}}]{Connors2022}%
  \BibitemOpen
  \bibfield  {author} {\bibinfo {author} {\bibfnamefont {E.~J.}\ \bibnamefont
  {Connors}}, \bibinfo {author} {\bibfnamefont {J.}~\bibnamefont {Nelson}},
  \bibinfo {author} {\bibfnamefont {L.~F.}\ \bibnamefont {Edge}},\ and\
  \bibinfo {author} {\bibfnamefont {J.~M.}\ \bibnamefont {Nichol}},\ }\href
  {https://doi.org/10.1038/s41467-022-28519-x} {\bibfield  {journal} {\bibinfo
  {journal} {Nature Communications}\ }\textbf {\bibinfo {volume} {13}},\
  \bibinfo {pages} {940} (\bibinfo {year} {2022})}\BibitemShut {NoStop}%
\end{thebibliography}%

\end{document}